\documentclass[aps,prb,twocolumn,superscriptaddress,showpacs,amsmath,amssymb]{revtex4-1}

\usepackage{amsfonts}
\usepackage{amssymb}
\usepackage{graphicx}
\usepackage{color}

\usepackage{bm}
\usepackage{braket}
\usepackage{gensymb}

\usepackage{hyperref}

\begin{document}

\title{Corner charge and bulk multipole moment in periodic systems}

\author{Haruki Watanabe} \email[]{hwatanabe@g.ecc.u-tokyo.ac.jp}
\author{Seishiro Ono}
\affiliation{Department of
Applied Physics, University of Tokyo, Tokyo 113-8656, Japan.}

\begin{abstract}
A formula for the corner charge in terms of the bulk quadrupole moment is derived for two-dimensional periodic systems. 
This is an analog of the formula for the surface charge density in terms of the bulk polarization.
In the presence of an $n$-fold rotation symmetry with $n=3$, $4$, and $6$, the quadrupole moment is quantized and is independent of the spread or shape of Wannier orbitals, depending only on the location of Wannier centers of filled bands.
In this case, our formula predicts the fractional part of the quadrupole moment purely from the bulk property.  The system can contain many-body interactions as long as the ground state is gapped and topologically trivial in the sense it is smoothly connected to a product state limit.  An extension of these results to three-dimensional systems is also discussed. In three dimensions, in general, even the fractional part of the corner charge is not fully predictable from the bulk perspective even in the presence of point group symmetry.
\end{abstract}

\maketitle

\section{Introduction}
The ``modern theory" of electric polarization~\cite{king-smith1993,vanderbilt1993,resta1994,resta2007,vanderbilt2018} succeeded in characterizing the surface charge of band insulators from the bulk polarization formulated in terms of the Berry phase of Bloch functions.  Recently, multipole insulators that feature fractional charges localized not on the surface but around hinges and corners of a slab attracted significant attention~\cite{benalcazarScience,benalcazar2017,song2017,ezawa2018,Imhof:2018aa,Peterson:2018aa,Serra-Garcia:2018aa,PhysRevB.98.045125,PhysRevB.98.081110,PhysRevB.98.201114,PhysRevLett.122.086804,He:2020aa,Peterson1114,PhysRevResearch.2.012067,PhysRevB.101.115115,PhysRevB.101.115140,PhysRevResearch.2.033029,PhysRevLett.124.036803,1908.00011,1909.05868,2001.07724,2002.08385,PhysRevX.9.031003,PhysRevResearch.2.012009,Hirayama_honeycomb}.  There have been several attempts~\cite{kang2018,wheeler2018} in extending the theory of polarization to the theory of the multipole moments that predicts the hinge charge density and the corner charge in terms of the bulk multipole moments.  However, the proposals~\cite{kang2018,wheeler2018} contain several fundamental issues associated with the periodicity of the boundary condition~\cite{ono2019}.  There are also several other recent proposals~\cite{LukaOnoWatanabe,Luka2020}: Ref.~\onlinecite{LukaOnoWatanabe} characterized the corner charge using the third Chern-Simons form for an adiabatic pumping process, but this framework requires a smooth interpolation of the quadrupole insulator to a trivial insulator. Ref.~\onlinecite{Luka2020} proposed a ``bulk-and-edge to corner'' correspondence, focusing on systems without rotation symmetry. However, in the presence of a rotation symmetry, the detailed information on the edge does not seem necessary.

Indeed, for two-dimensional band insulators with a rotation symmetry, formulas predicting corner charges in terms of the rotation representations of Bloch functions have been developed~\cite{benalcazar2018,PhysRevResearch.1.033074,Shiozaki2019}.  
This ``symmetry-indicator"~\cite{po2017,khalaf2018} type approach fulfills the criterion of describing the corner charge purely based on the bulk property of band insulators. However, there remain three unsatisfactory points: (i) The relation to multipole moments is unclear. (ii) The formulas are incomplete in the sense that it is not always possible to predict the corner charge based on the rotation representations alone. Examples are given in Ref.~\onlinecite{PhysRevResearch.1.033074}. (iii) Rotation representations of Bloch functions are fundamentally affected by whether or not the spin-orbit coupling is taken into account and the time-reversal symmetry is assumed.  Thus, in this approach, formulas must be derived separately for each setup.  
Both of the previous works~\cite{benalcazar2018,PhysRevResearch.1.033074,Shiozaki2019} assumed the time-reversal symmetry, and the more general case remains an open problem.

In this work, we develop a theory that improves all of these points and establish a ``bulk-corner" correspondence.  We first formulate the corner charge of two-dimensional periodic systems in general in terms of the bulk quadrupole moment [Eq.~\eqref{cornercharge}] in a way the analogy to the modern-theory formula of the surface charge [Eq.~\eqref{surfacecharge}] is evident.  We then add the $n$-fold rotation symmetry ($n=3$, $4$, and $6$) to the problem and show that the general formula [Eq.~\eqref{cornercharge}] reduces to a simpler one [Eq.~\eqref{rotationformula}] formulated in terms of the U(1) charges localized at each Wyckoff position~\cite{ITC}. This formula predicts the fractional part of the quantized corner charge based on bulk topological invariants protected by the particle number conservation, the rotation symmetry, and the lattice translation symmetry.  The formula works when the bulk system is charge neutral and polarization free, and applies even to interacting systems. The assumption of the formula, in addition to the symmetries and the lack of the charge density and the bulk polarization, is that the ground state is topologically trivial in the sense it is adiabatically connected to an atomic limit~\cite{PhysRevB.99.125122}.  We also extend these results to three-dimensional systems, deriving formulas of the hinge charge density [Eq.~\eqref{hingecharge}] and the corner charge [Eq.~\eqref{cornercharge3D}] in terms of the quadrupole moment and the octupole moment.

\section{Setting and definitions}
\label{sec:setting}
In this section we summarize the setup of our study. We consider U(1) invariant systems with a lattice translation symmetry in $d$ spatial dimensions. In this work we are interested in $d=2$ and $3$. The reciprocal lattice vectors $\bm{b}_i$'s are defined by $\bm{a}_i\cdot\bm{b}_j=\delta_{ij}$ (without $2\pi$) for given primitive lattice vectors $\bm{a}_i$ ($i,j=1,\dots,d$). 

The assumed U(1) symmetry implies the conservation of the total U(1) charge of the system $\hat{Q}=\int_{\mathbb{R}^d} d^dr \hat{\rho}(\bm{r})$.
\footnote{In the oblique coordinate system $\bm{r}=\sum_{i=1}^dr_i\bm{a}_i$,
\begin{align}
\int_{\mathbb{R}^d} d^dr\equiv v\int_{-\infty}^{+\infty} dr_1\cdots \int_{-\infty}^{+\infty} d r_d,
\end{align}
where $v$ is the volume (or the area) of a unit cell.}
We can unambiguously define the total charge density of the system $\rho_{\text{tot}}^{\text{(bulk)}}(\bm{r})$  as the ground-state expectation value of the charge density operator $\hat{\rho}(\bm{r})$. The translation invariance implies 
\begin{align}
\rho_{\text{tot}}^{\text{(bulk)}}(\bm{r}-\textstyle\sum_{i=1}^dn_i\bm{a}_i)=\rho_{\text{tot}}^{\text{(bulk)}}(\bm{r})
\end{align}
for any $n_i\in\mathbb{Z}$. 

We decompose $\rho_{\text{tot}}^{\text{(bulk)}}(\bm{r})$ into the sum of a local charge density $\rho_0(\bm{r})$:
\begin{align}
\rho_{\text{tot}}^{\text{(bulk)}}(\bm{r})=\sum_{n_i\in\mathbb{Z}}\rho_0(\bm{r}-\textstyle\sum_{i=1}^dn_i\bm{a}_i).
\label{totalcharge}
\end{align}
The local charge density $\rho_0(\bm{r})$ is composed of electronic and ionic orbitals labeled by $\alpha$ and $\beta$: 
\begin{equation}
\rho_0(\bm{r})=\sum_{\alpha=1}^{\nu_{\text{el}}} \rho_{0\alpha}^{\text{(el)}}(\bm{r})+\sum_{\beta=1}^{\nu_{\text{ion}}} \rho_{0\beta}^{\text{(ion)}}(\bm{r}).
\end{equation}
Both $\rho_{0\alpha}^{\text{(el)}}(\bm{r})$ and $\rho_{0\beta}^{\text{(ion)}}(\bm{r})$ are assumed to be exponentially localized and are normalized to an integer multiple of the unit charge $e$ ($>0$):
\begin{align}
\int_{\mathbb{R}^d} d^dr\rho_{0\alpha}^{\text{(el)}}(\bm{r})=-e,\quad \int_{\mathbb{R}^d} d^dr\rho_{0\beta}^{\text{(ion)}}(\bm{r})=m_{\beta}e.
\end{align}
The charge neutrality imposes the condition $\nu_{\text{el}}=\sum_{\beta=1}^{\nu_{\text{ion}}}m_{\beta}$ on the number of orbitals $\nu_{\text{el}}$, $\nu_{\text{ion}}$ per unit cell and ionic charges $m_{\beta}e$.  For band insulators, $\rho_{0\alpha}^{\text{(el)}}(\bm{r})$'s are constructed as Wannier orbitals~\cite{vanderbilt2018} of filled bands (see Sec.~\ref{sec:BI} for more details)~\cite{marzari1997,RevModPhys.84.1419}, while $\rho_{0\beta}^{\text{(ion)}}(\bm{r})$'s are usually simply given by atomic orbitals.   Our general formulation treats electrons and ions on equal footing.

Note that the correspondence between $\rho_{\text{tot}}^{\text{(bulk)}}(\bm{r})$ and $\rho_0(\bm{r})$ is \emph{one-to-many}. That is, there are multiple possible choices of $\rho_0(\bm{r})$ that give the same bulk charge density $\rho_{\text{tot}}^{\text{(bulk)}}(\bm{r})$, and there is no unique way of determining $\rho_0(\bm{r})$ based on $\rho_{\text{tot}}^{\text{(bulk)}}(\bm{r})$. Here we proceed with a given $\rho_0(\bm{r})$, paying attention to its ambiguity.

For a function $F(\bm{r})$ of $\bm{r}$, we denote by $\langle F(\bm{r})\rangle_0$ the spatial average of $F(\bm{r})$ with respect to $\rho_0(\bm{r})$:
\begin{equation}
\langle F(\bm{r})\rangle_0\equiv \int_{\mathbb{R}^d} d^dr\rho_0(\bm{r})F(\bm{r}).
\label{fav}
\end{equation}
The charge neutrality implies 
\begin{equation}
\langle 1\rangle_0=0.
\end{equation}

\begin{figure}[t]
\begin{center}
\includegraphics[width=\columnwidth]{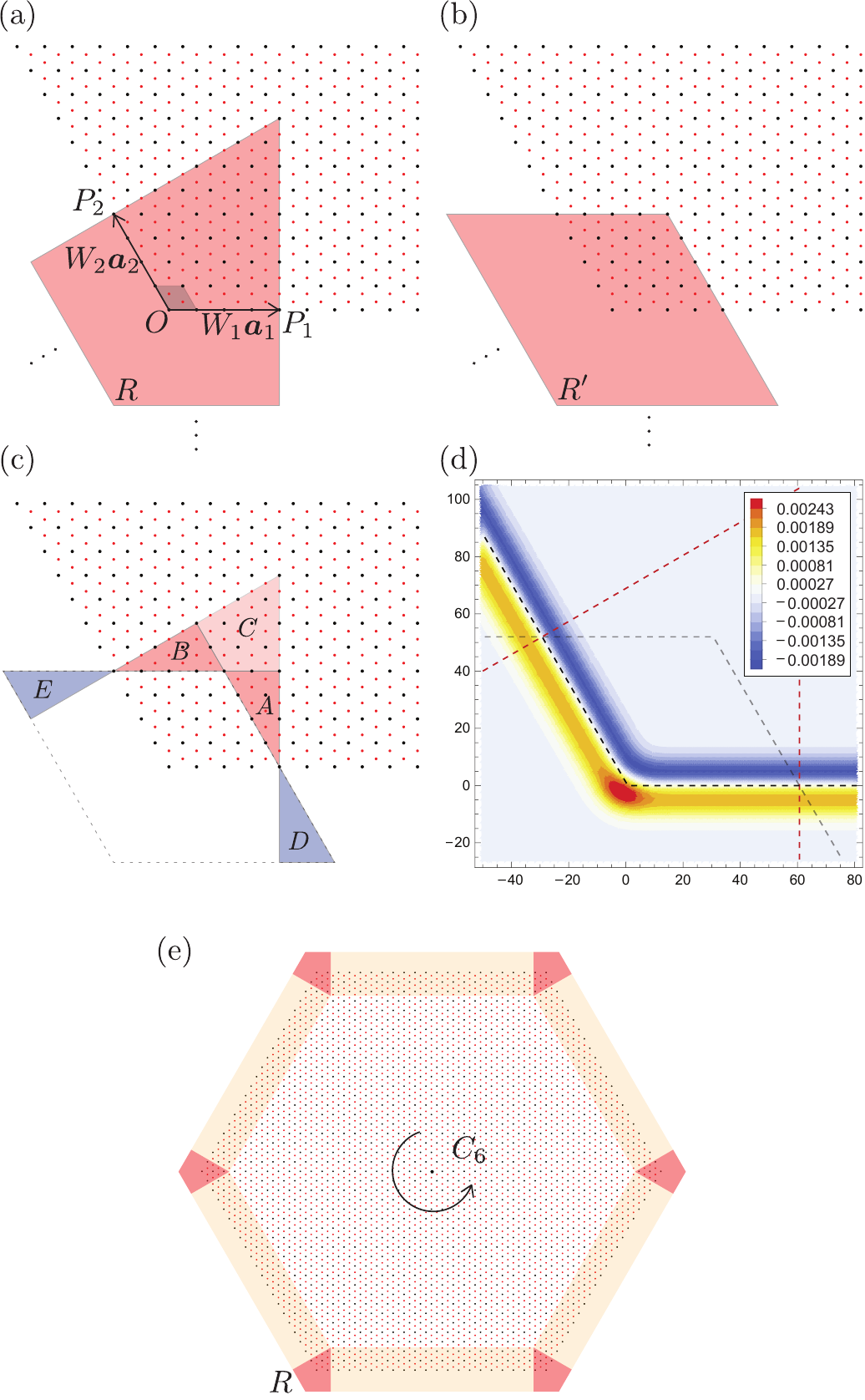}
\caption{\label{figR} (a) The region $R$ used in the definition of the corner charge $Q_c$ in $d=2$. A unit cell is shown by a gray parallelogram.  (b) The region $R'$. Each boundary is parallel to either $\bm{a}_1$ or $\bm{a}_2$. (c) The difference of $R$ and $R'$. (d) The contour plot of the coarse-grained density $\tilde{\rho}_{\text{tot}}(\bm{r})$ in Eq.~\eqref{totalchargec2} for honeycomb lattice in Fig.~\ref{figC6C3} (c) for $a_1=a_2=1$, $\lambda=5$, and $e=1$. Some ions are shifted so that the bulk polarization vanishes [see Fig.~\ref{figC6C3} (c) for the detail]. 
Both $\rho_{0\alpha}^{\text{(el)}}(\bm{r})$ and $\rho_{0\beta}^{\text{(ion)}}(\bm{r})$ are chosen to be delta functions in this plot.
(e) An example of finite systems with six-fold rotation symmetry. }
\end{center}
\end{figure}

Let us introduce a corner by restricting the sum in Eq.~\eqref{totalcharge} to  $n_i\geq0$ for every $i=1,\dots,d$:
\begin{equation}
\rho_{\text{tot}}(\bm{r})\equiv\sum_{n_i\geq0}\rho_0(\bm{r}-\textstyle\sum_{i=1}^dn_i\bm{a}_i).
\label{totalchargec}
\end{equation}
The corner is defined by surfaces normal to $\bm{b}_1, \dots,\bm{b}_d$.  Different choices of $\rho_0(\bm{r})$ for the same $\rho_{\text{tot}}^{\text{(bulk)}}(\bm{r})$ result in different types of terminations of the surfaces.
In reality, the charge density near the surface may be reconstructed, but for now we neglect such an effect. We will revisit this point later.

Given $\rho_{\text{tot}}(\bm{r})$ in Eq.~\eqref{totalchargec}, one can compute the total charge $Q_R$ in the region $R$ illustrated in Fig.~\ref{figR} (a).  The edges of $R$ intersect with the boundary of the system at $P_i$ ($i=1,\ldots,d$) [see Fig.~\ref{figR} (a)], and the vector connecting the origin $O$ to $P_i$ is $\bm{a}_iW_i$. To properly characterize the corner charge, the edges of the region $R$ must be orthogonal to the boundary of the system.  An example of invalid choice is the region $R'$ in Fig.~\ref{figR} (b).  The distinction between $R$ and $R'$ remains important even when the surface and hinge charge density vanishes (see Sec.~\ref{sec:2Dformula} for an example).  Mathematically, the region $R$ is specified by the conditions
\begin{align}
&(\bm{r}-W_i\bm{a}_i)\cdot\bm{a}_i=a_i^2\Big[(r_i-W_i)-\sum_{j\neq i}c_{ij}r_j\Big]<0
\label{RR}
\end{align}
for $i=1,\ldots,d$, where
\begin{equation}
c_{ij}\equiv-\frac{\bm{a}_i\cdot\bm{a}_j}{a_i^2}.
\label{cij}
\end{equation}

The microscopic charge density $\rho_0(\bm{r})$ may be highly oscillating at a scale equal to or even much smaller than the lattice constant $a_i$, and the total charge in $R$ depends sensitively on the location of the boundary of $R$. To avoid such subtlety, we perform coarse graining of the charge density by a convolution integral with the Gaussian function $G(\bm{r})\equiv (2\pi\lambda^2)^{-D/2}e^{-\frac{|\bm{r}|^2}{2\lambda^2}}$ (see Sec.~6.6 of \onlinecite{jackson2007classical}):
\begin{align}
&\tilde{\rho}_{\text{tot}}(\bm{r})\equiv\int_{\mathbb{R}^d}d^dr' G(\bm{r}-\bm{r}')\rho_{\text{tot}}(\bm{r}'),\label{totalchargec2}\\
&\tilde{\rho}_0(\bm{r})\equiv\int_{\mathbb{R}^d}d^dr' G(\bm{r}-\bm{r}')\rho_0(\bm{r}').
\end{align}
Here, the parameter $\lambda$ ($\gg a_i$) specifies the scale after coarse graining, and the width $W_i$ [see Fig.~\ref{figR} (a)] is assumed to be much greater than $\lambda$. In contrast to the microscopic density $\rho_0(\bm{r})$, the coarse-grained one $\tilde{\rho}_0(\bm{r})$ is a slowly varying, smooth function of $\bm{r}$.  This gives a legitimate, stable definition of the total charge in the region $R$:
\begin{equation}
Q_R\equiv\int_{R}d^dr\tilde{\rho}_{\text{tot}}(\bm{r}).
\label{defQR}
\end{equation}

Finally, let us define $\langle F(\bm{r})\rangle_{\tilde{0}}$ as the spatial average of $F(\bm{r})$ with respect to the coarse-grained density $\tilde{\rho}_0(\bm{r})$:
\begin{equation}
\langle F(\bm{r})\rangle_{\tilde{0}}\equiv \int_{\mathbb{R}^d} d^dr\tilde{\rho}_0(\bm{r})F(\bm{r}).
\label{fav2}
\end{equation}
Using $G(\bm{r})=G(-\bm{r})$, we see that
\begin{equation}
\langle F(\bm{r})\rangle_{\tilde{0}}=\int_{\mathbb{R}^d} d^dr\int_{\mathbb{R}^d}d^dr' G(\bm{r}-\bm{r}')\rho_0(\bm{r}')F(\bm{r})=\langle \tilde{F}(\bm{r})\rangle_0,
\end{equation}
where 
\begin{align}
\tilde{F}(\bm{r})\equiv\int_{\mathbb{R}^d}d^dr' G(\bm{r}-\bm{r}')F(\bm{r}').
\end{align}
It can be readily verified that $\langle 1\rangle_{\tilde{0}}=\langle 1\rangle_0=0$ and $\langle r_i\rangle_{\tilde{0}}=\langle r_i\rangle_0$. However, in general, $\langle F(\bm{r})\rangle_{\tilde{0}}$ and $\langle F(\bm{r})\rangle_0$ do not agree. 
For example, in the case of $F(\bm{r})=x^3$, $\langle x^3\rangle_{\tilde{0}}-\langle x^3\rangle_0=3\lambda^2\langle x\rangle_0$.
Nonetheless, we will see that the part of quadrupole moments and octupole moments relevant for hinge and corner charges exhibits the ``coarse-graining invariance"
\begin{align}
F(\bm{r})=\tilde{F}(\bm{r})
\label{invariance}
\end{align}
so that $\langle F(\bm{r})\rangle_0=\langle F(\bm{r})\rangle_{\tilde{0}}$.

\section{Two dimensions}
\label{sec:result}
In this section we present our results for two-dimensional systems. 

\subsection{Formula for corner charge in two dimensions}
\label{sec:2Dformula}
Without loss of generality, the primitive lattice vectors $\bm{a}_i$  ($i=1,2$) can be set 
\begin{equation}
\bm{a}_1=a_1(1,0),\,\,\,\,\bm{a}_2=a_2(\cos\theta,\sin\theta)\,\,\,\,(0<\theta<\pi).
\label{QR1}
\end{equation}
For $n$-fold rotation symmetric systems ($n=3$, $4$, or $6$), we set $a_1=a_2$ and $\theta=\pi-(2\pi/n)$ so that $\bm{a}_2$ is mapped to $-\bm{a}_1$ under $2\pi/n$-rotation. Thus the square lattice ($n=4$) and the hexagonal lattice ($n=6$) correspond to $\theta=\pi/2$ and $\theta=2\pi/3$, respectively.

\subsubsection{Surface charge}
When the bulk polarization does not vanish, $Q_R$ in Eq.~\eqref{defQR} is dominated by the contributions from the surface:
\begin{equation}
Q_R=W_1\sigma_2+W_2\sigma_1+O(1).
\end{equation}
The surface charge density $\sigma_i$ ($\sigma_1$ is per length $a_2$ and $\sigma_2$ is per length $a_1$) is given by the bulk polarization~\cite{king-smith1993,vanderbilt1993,resta2007,vanderbilt2018} 
\begin{align}
&\sigma_i=-\langle P_i(\bm{r})\rangle_0=-\langle P_i(\bm{r})\rangle_{\tilde{0}}\label{surfacecharge},\\
&P_i(\bm{r})\equiv\bm{b}_i\cdot \bm{r}=r_i.
\end{align}
For band insulators, the electronic contribution to $\langle P_i(\bm{r})\rangle_0$ is given by the sum of the Berry phase of filled bands [see Eq.~\eqref{Berry} below]~\cite{king-smith1993,vanderbilt1993} .

\subsubsection{Corner charge}
When the bulk polarization vanishes, $Q_R$ measures the charge bound to the corner
\begin{equation}
Q_R=Q_c.
\end{equation}
The first main result of this work is the following formula for the corner charge:
\begin{align}
Q_c&=\langle Q_{12}(\bm{r})\rangle_0=\langle Q_{12}(\bm{r})\rangle_{\tilde{0}},\label{cornercharge}\\
Q_{12}(\bm{r})&\equiv(\bm{b}_1\cdot\bm{r})(\bm{b}_2\cdot\bm{r})\notag\\
&\quad\quad\quad+\frac{\bm{a}_2\cdot\bm{a}_1}{2a_2^2}(\bm{b}_1\cdot\bm{r})^2+\frac{\bm{a}_1\cdot\bm{a}_2}{2a_1^2}(\bm{b}_2\cdot\bm{r})^2\notag\\
&=r_1r_2+\frac{1}{2}\cos\theta\left(\frac{a_1}{a_2} r_1^2+\frac{a_2}{a_1}r_2^2\right)\notag\\
&=\frac{1}{a_1a_2}\left(\frac{x^2-y^2}{2}\cos\theta+ x y\sin\theta\right).\label{cornercharge222}
\end{align}
The second line of Eq.~\eqref{cornercharge222} is for the oblique coordinate $\bm{r}=r_1\bm{a}_1+r_2\bm{a}_2$ and the third line is for the Cartesian coordinate $\bm{r}=(x,y)$. The quantity $\langle Q_{12}(\bm{r})\rangle_0$ can be interpreted as the bulk quadrupole moment.  The same result for the square lattice and the cubic lattice has been derived before in Ref.~\onlinecite{benalcazar2017}, and our formula extends it to arbitrary lattices.  Note that both $P_i(\bm{r})$ and $Q_{12}(\bm{r})$ satisfy the coarse-graining invariance \eqref{invariance}.  Furthermore, they are independent of the choice of the origin when lower multipoles vanish. These observations support the validity of our results.
We present the derivation of Eqs.~\eqref{surfacecharge} and \eqref{cornercharge} in Sec.~\ref{sec:derivation}. As shown there, the total charge in $R'$, when $\langle P_i(\bm{r})\rangle_0=0$, is given by
\begin{equation}
Q_{R'}=\langle(\bm{b}_1\cdot\bm{r})(\bm{b}_2\cdot\bm{r})\rangle_0.
\end{equation}
This is only a part of $Q_c$ in Eq.~\eqref{cornercharge}.

The origin of the discrepancy between $Q_R$ and $Q_{R'}$ can be understood by focusing on the profile of the charge density near the surface. In general, we have [see Fig.~\ref{figR} (c)]
\begin{align}
Q_R=Q_{R'}+Q_A+Q_B+Q_C-Q_D-Q_E.
\label{QRQRp}
\end{align}
Even when the bulk polarization vanishes and no net surface charge is expected, the charge density profile $\tilde{\rho}_{\text{tot}}(\bm{r})$ may not completely vanish near the surface, showing some spatial dependence as in Fig.~\ref{figR} (d). 
In the example of the honeycomb lattice in Fig.~\ref{figR},  $Q_A$ and $Q_B$ are negative and $Q_D$ and $Q_E$ are positive. $Q_C=0$ due to the charge neutrality in the bulk. Consequently, $Q_{R'}$ should be larger than $Q_{R}$, and we indeed find $Q_{R'}=5e/9$ and $Q_R=e/3$ in this example. The modulation of $\tilde{\rho}_{\text{tot}}(\bm{r})$ does not affect $Q_R$ because the net surface charge in the light-yellow region in Fig.~\ref{figR} (e) vanishes owing to the assumed orthogonality of the boundary of $R$ to the system. This is why $Q_R$ gives the correct quantized value, consistent with the previous study~\cite{benalcazar2018}.

Note that the formulas in Eqs.~\eqref{surfacecharge} and \eqref{cornercharge} sensitively depend on the detailed shape of $\rho_0(\bm{r})$. In particular, the value of $\langle Q_{12}(\bm{r})\rangle_0$ can be smoothly changed without affecting the bulk polarization $\langle P_i(\bm{r})\rangle_0$.  This observation implies that, in general,  $\langle Q_{12}(\bm{r})\rangle_0$ for band insulators is ill-defined because of the gauge ambiguity in forming the Wannier orbitals that may affect their shape.

\begin{figure}
\begin{center}
\includegraphics[width=1\columnwidth]{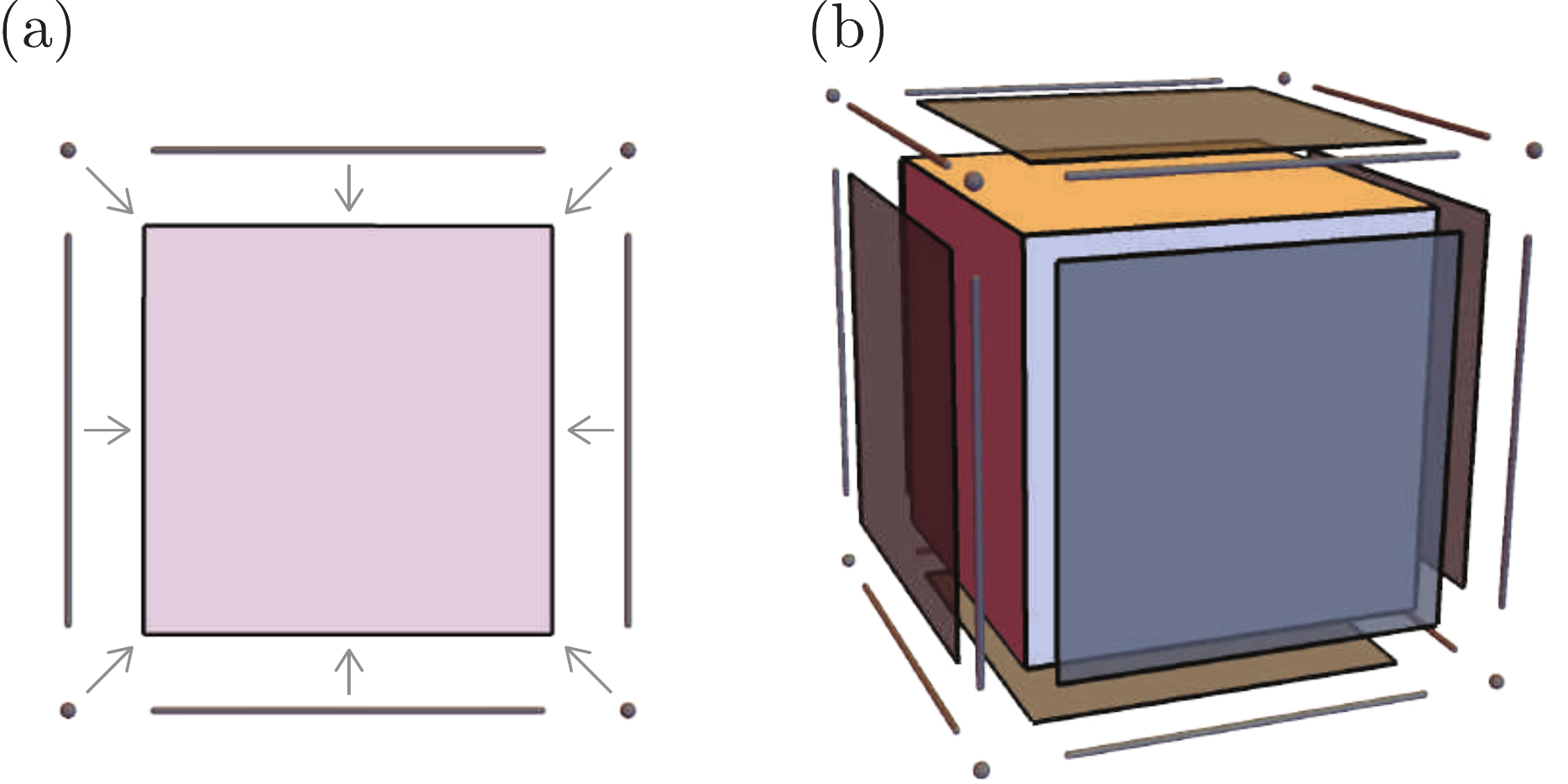}
\caption{\label{figdecoration} Decoration of surfaces, hinges, and corners by lower dimensional objects for (a) two-dimensional and (b) three-dimensional systems.}
\end{center}
\end{figure}

Furthermore, the above formula is designed for the particular termination of the system specified above. The corner charge as well as the surface charge density can be altered by decoration of the surface with lower dimensional objects with a nonzero charge or polarization as illustrated in Fig.~\ref{figdecoration}.

\subsubsection{Rotation symmetry}
The issues mentioned just now can be suppressed in the presence of a rotation symmetry.  Crucially,  $\langle Q_{12}(\bm{r})\rangle_0$ depends only on the location of the ``Wannier center," and does not depend on the detailed shape of $\rho_0(\bm{r})$ as long as the rotation symmetry is properly implemented.  This means that $\langle Q_{12}(\bm{r})\rangle_0$ becomes well defined for band insulators under the rotation symmetry.
Moreover, $\langle Q_{12}(\bm{r})\rangle_0$ is quantized  and is robust against smooth deformation such as the surface reconstruction.  The integer part of the corner charge can still be altered by symmetrically attaching a charged object to each corner [see  Fig.~\ref{figdecoration} (a)], and cannot be predicted only from the bulk property.

\begin{figure*}
\begin{center}
\includegraphics[width=0.85\textwidth]{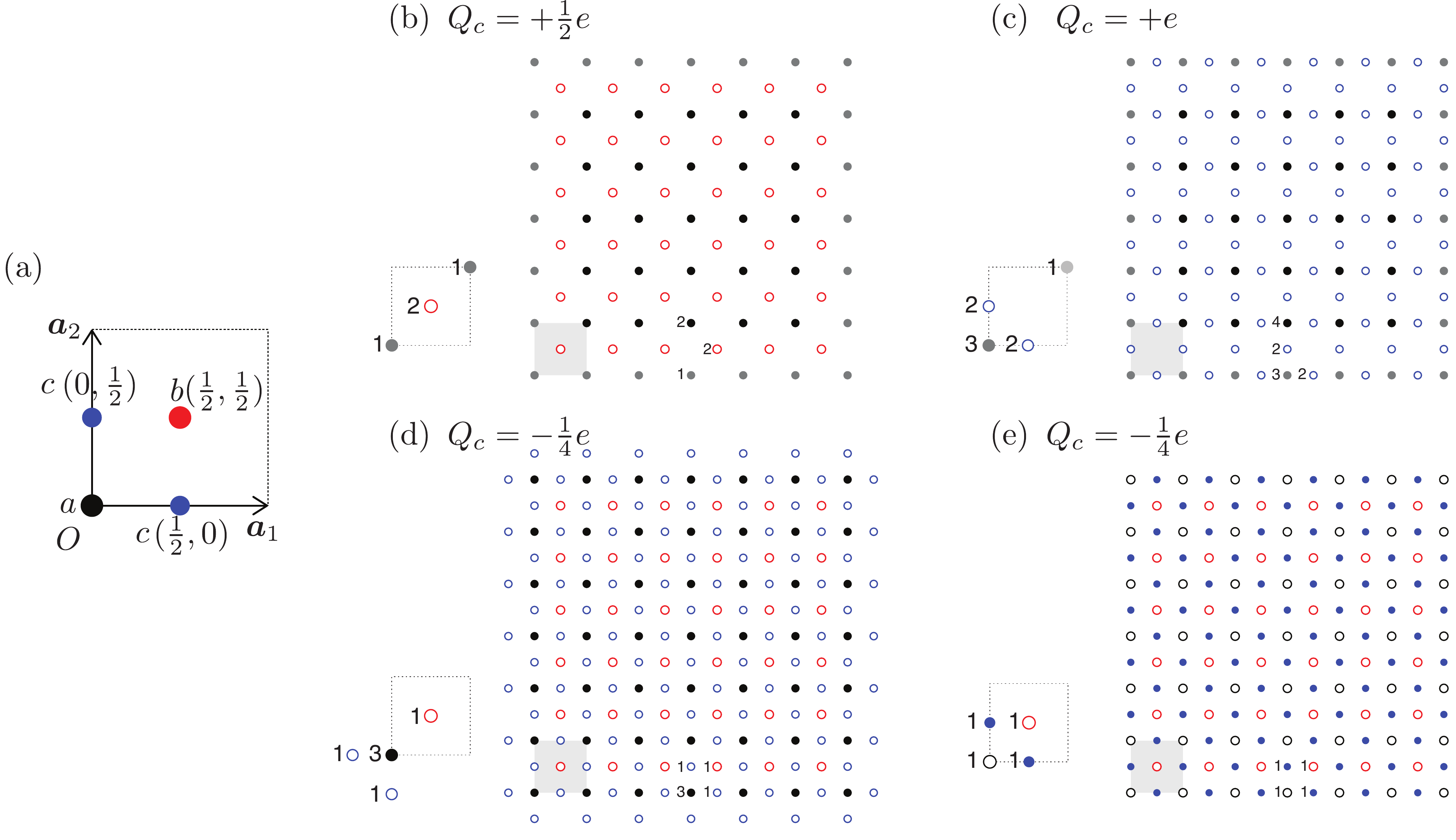}
\caption{\label{figC4} (a): Wyckoff positions for $C_4$ symmetry. (b)-(e): Examples of $C_4$-symmetric systems without bulk polarization. Electronic (ionic) orbitals are represented by open (solid) circles. Colors correspond to Wyckoff positions in the panel  (a). The repetition unit $\rho_0(\bm{r})$ is shown at the top-left of each panel. Integers next to circles represent the number of orbitals at the location.}
\end{center}
\end{figure*}

\begin{figure*}
\begin{center}
\includegraphics[width=0.85\textwidth]{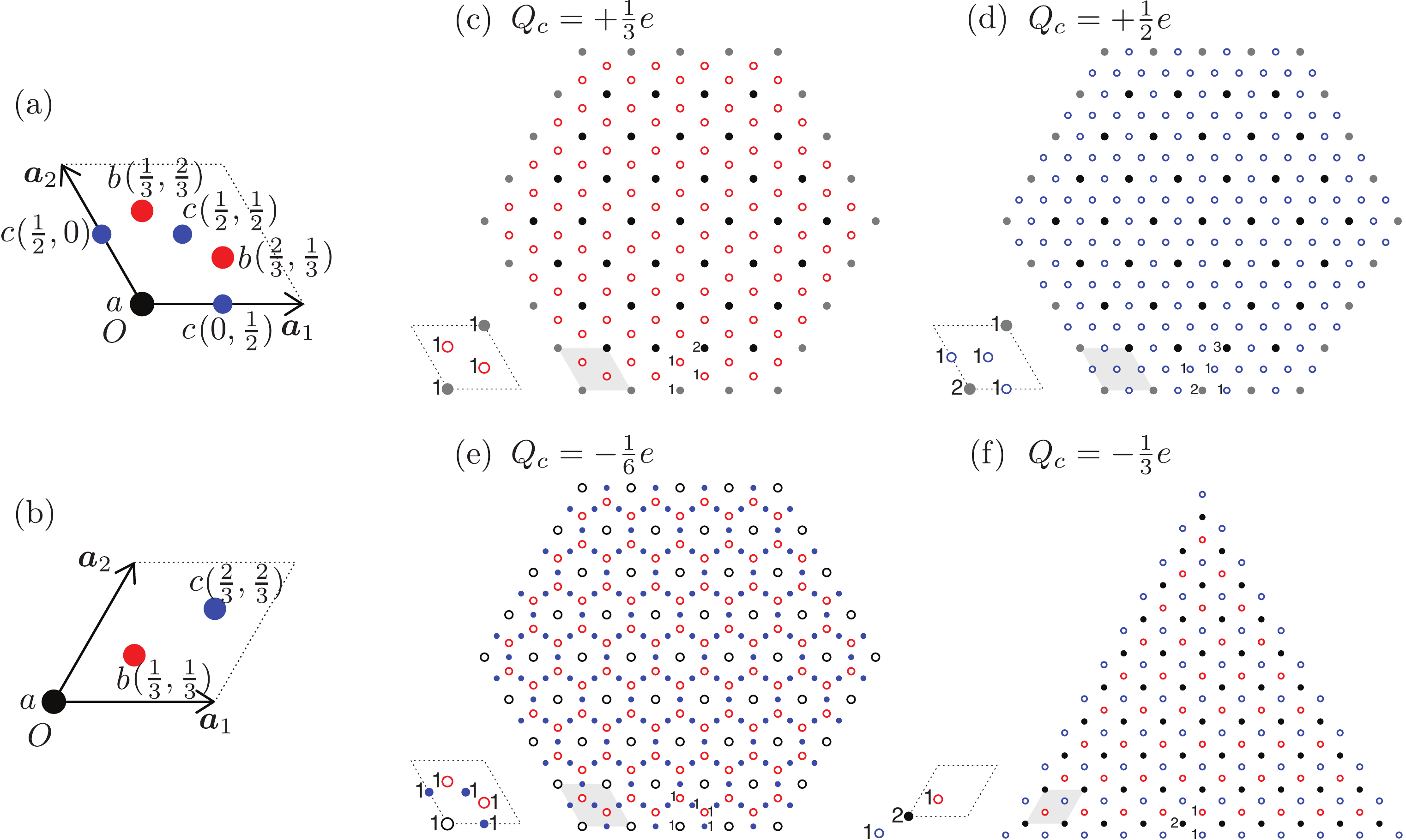}
\caption{\label{figC6C3} (a), (b): Wyckoff positions for $C_6$ symmetry [(a)] and for $C_3$ symmetry [(b)]. Coordinates are given in oblique systems $\bm{r}=r_1\bm{a}_1+r_2\bm{a}_2$. (c)-(f) Examples of $C_6$-symmetric [(c)-(e)] systems and a $C_3$-symmetric system [(f)] without bulk polarization. Electronic (ionic) orbitals are represented by open (solid) circles. Colors correspond to Wyckoff positions in the panel (a) or (b). The repetition unit $\rho_0(\bm{r})$ is shown at the bottom-left of each panel. Integers next to circles represent the number of orbitals at the location.}
\end{center}
\end{figure*}

The second main result of this work is the following formula of the fractional part of $\langle Q_{12}(\bm{r})\rangle_0$ in terms of the occupation of each Wyckoff position under $n$-fold rotation symmetry:
\begin{eqnarray}
\langle Q_{12}(\bm{r})\rangle_0=
\begin{cases}
\frac{1}{4}q_a=\frac{1}{4}q_b\mod e& (n=4)\\
\frac{1}{6}q_a=\frac{2}{3}q_b+\frac{1}{2}q_c\mod e& (n=6)\\
\frac{1}{3}q_a=\frac{1}{3}q_b=\frac{1}{3}q_c\mod e& (n=3),
\end{cases}
\label{rotationformula}
\end{eqnarray}
where $q_w$ is the total U(1) charge per site belonging to the Wyckoff position $w$ ($=a$, $b$, $c$). See Figs.~\ref{figC4} (a) and \ref{figC6C3} (a), (b) for the illustration of the Wyckoff position. Relations such as $q_a=q_b$ mod $4e$ for $n=4$ and $q_a=q_b=q_c$ mod $3e$ for $n=3$ follow by the requirement of the charge neutrality and the vanishing polarization. These relations make our formula \eqref{rotationformula} independent of the choice of the origin of the unit cell. We present the derivation of Eq.~\eqref{rotationformula} in Sec.~\ref{sec:quantization}.  The above formula for $n=4$ in terms of $q_a$ was derived before in Ref.~\onlinecite{PhysRevB.98.081110} for band insulators when ionic positions are restricted to Wyckoff position $a$. 

To explain how to use the formula, let us consider the case of $n=6$. The Wyckoff positions $w=a$, $b$, and $c$, respectively, correspond to the triangular lattice, the honeycomb lattice, and the kagome lattice. When an electronic Wannier orbital sits at $w=b$ and two ions sit at $w=a$ [see Fig.~\ref{figC6C3} (c)], then one has $q_b=-e$ and $q_a=2e$ so that $Q_c=\langle Q_{12}(\bm{r})\rangle_0=e/3$ (mod $e$).  Several other examples are shown in Figs.~\ref{figC4} and \ref{figC6C3}. 

The quantities appearing in the right-hand side of Eq.~\eqref{rotationformula} are purely bulk topological invariants in the sense that they can be fully determined by the (many-body) ground state $|\Phi_0\rangle$ for systems under the periodic boundary condition or for the infinite system without boundaries and that they are robust against smooth deformation~\cite{PhysRevB.99.125122}. For example, any topologically trivial band insulator can be almost uniquely decomposed into a stack of atomic insulators and $q_{w}$ represents the coefficients of the superposition.  The undetermined part of $q_w$ arises from the ``lattice homotopy" equivalence~\cite{PhysRevLett.119.127202,PhysRevB.99.125122}: Wyckoff positions with free parameters can be smoothly reduced to some of the special positions, making $q_w$ well defined only modulo some integers. For example, in the case of $n=4$, $\bm{r}_d^{(\ell)}$ in Eq.~\eqref{4d} reduces to $\bm{r}_a^{(\ell)}$,  $\bm{r}_b^{(\ell)}$, and $\bm{r}_c^{(\ell)}$  in Eqs.~\eqref{4a}--\eqref{4c} by setting $(r_1, r_2)=(0,0)$, $(1/2,1/2)$, and $(1/2,0)$, respectively. (Neglect the integer part and set $k=k'=0$ for the purpose of the discussion on the bulk charge distribution here.) In general, if the site symmetry of the Wyckoff position is $m_w$-fold rotation, then $q_w$ is defined modulo $m_we$. However, such an ambiguity does not affect the fractional part of Eq.~\eqref{rotationformula} in two dimensions. In contrast, we will see in Sec.~\ref{threedim} that the ambiguity of $q_w$ affects even the fractional part of $\langle Q_{123}(\bm{r})\rangle_0$ in three dimensions.

Given a many-body Hamiltonian $\hat{H}$ of the system, the easiest way of determining the U(1) charge $q_w$ of a Wyckoff position $\bm{r}_w^{(\ell)}$ would be via the computation of the total U(1) charge of the ground state $|\Phi_0\rangle$ under an open boundary condition designed for each Wyckoff position.  When the site symmetry of the Wyckoff position $\bm{r}_w^{(\ell)}$ is $m_w$-fold rotation symmetry, we prepare a finite-size system with the $m_w$-fold rotation symmetry about the rotation axis $\bm{r}=\bm{r}_w^{(\ell)}$.  The linear dimension of the system must be sufficiently larger than the correlation length.  Then the physical properties of this finite system around the rotation axis should be identical to those for the infinite system. 
Because of the assumed rotation symmetry, the total U(1) charge $Q_w$ of the entire system under the open boundary condition must coincide with the U(1) charge $q_w$ bound to the rotation axis, modulo $m_we$:
\begin{equation}
Q_w=q_w \mod m_we.
\label{Qwqw}
\end{equation}
As an example, let us consider the charge configuration illustrated in Fig.~\ref{figC4} (b),  (c),  (d), and  (e). They have the four-fold rotation symmetry ($m_{w=a}=4$) about the rotation axis $\bm{r}=\bm{r}_{w=a}^{(\ell)}$ at the center.  By the direct calculation, we find $Q_{w=a}=2e$, $4e$, $-e$, $-e$ and $q_{w=a}=2e$, $4e$, $3e$, $-e$, respectively, for these panels, confirming the relation in Eq.~\eqref{Qwqw}. 

\subsection{Derivation of the formula for two dimensions}
\label{sec:derivation}
Here we present the derivation of our formula for $\sigma_i$ and $Q_c$ in Eqs.~\eqref{surfacecharge} and \eqref{cornercharge}.

As noted before, the coarse-grained density $\tilde{\rho}_0(\bm{r})$ is a smooth, slowly varying function of $\bm{r}=r_1\bm{a}_1+r_2\bm{a}_2$.
This makes it possible to approximate the summation in the total charge density by an integral:
\begin{align}
\tilde{\rho}_{\text{tot}}(\bm{r})&\equiv\sum_{n_1,n_2\geq0}\tilde{\rho}_0(\bm{r}-n_1\bm{a}_1-n_2\bm{a}_2)\notag\\
&\simeq\int_{-\frac{1}{2}}^{+\infty} dr_1' \int_{-\frac{1}{2}}^{+\infty} dr_2'\,\tilde{\rho}_0(\bm{r}-r_1'\bm{a}_1-r_2'\bm{a}_2)\notag\\
&=\int_{-\infty}^{r_1+\frac{1}{2}} dr_1' \int_{-\infty}^{r_2+\frac{1}{2}} dr_2'\,\tilde{\rho}_0(r_1'\bm{a}_1+r_2'\bm{a}_2).
\label{chargedensity}
\end{align}
(The $1/2$'s appear here as a result of the midpoint prescription and does not affect the final result.) The last expression implies, among other things, that $\tilde{\rho}_{\text{tot}}(\bm{r})$ deep inside the bulk ($r_1,r_2\gg\lambda$) vanishes due to the charge neutrality,
\begin{align}
\tilde{\rho}_{\text{tot}}(\bm{r})&\simeq\int_{-\infty}^{+\infty} dr_1' \int_{-\infty}^{+\infty} dr_2'\,\tilde{\rho}_0(\bm{r}')=0.
\end{align}

We derive the expression for $Q_R$ via Eq.~\eqref{QRQRp}. We compute $Q_{R'}$ first and then take into account the difference of $R$ and $R'$.  We have
\begin{align}
&Q_{R'}\equiv \int_{R'}d^2r'\,\tilde{\rho}_{\text{tot}}(\bm{r}')=v\int_{-\infty}^{W_1}dr_1'\int_{-\infty}^{W_2}dr_2'\tilde{\rho}_{\text{tot}}(\bm{r}')\notag\\
&=v\int_{-\infty}^{W_1}dr_1'\int_{-\infty}^{W_2}dr_2'\int_{-\infty}^{r_1'+\frac{1}{2}} dr_1 \int_{-\infty}^{r_2'+\frac{1}{2}} dr_2\,\tilde{\rho}_0(\bm{r})\notag\\
&=v\int_{-\infty}^{W_1+\frac{1}{2}}dr_1\int_{-\infty}^{W_2+\frac{1}{2}}dr_2\int_{r_1-\frac{1}{2}}^{W_1} dr_1' \int_{r_2-\frac{1}{2}}^{W_2} dr_2'\,\tilde{\rho}_0(\bm{r})\notag\\
&=\left\langle\int_{r_1-\frac{1}{2}}^{W_1} dr_1' \int_{r_2-\frac{1}{2}}^{W_2} dr_2'\,1\right\rangle_{\tilde{0}}\notag\\
&=\left\langle (W_1-r_1+\tfrac{1}{2})(W_2-r_2+\tfrac{1}{2})\right\rangle_{\tilde{0}}
\label{QRp}
\end{align}
In going to the fourth line, we approximated $\int_{-\infty}^{W_i+\frac{1}{2}}dr_i$ by $\int_{-\infty}^{+\infty}dr_i$, which can be verified for a sufficiently large $W_i$.  

Similarly, relying on the fact that $\tilde{\rho}_{\text{tot}}(\bm{r})$ becomes independent of $r_1$ when $r_1>W_1\gg\lambda$, we have
\begin{align}
&Q_A-Q_D=v\int_{-\infty}^{W_2}dr_2'\int_{W_1}^{W_1+c_{12}r_2'}dr_1'\tilde{\rho}_{\text{tot}}(\bm{r}')\notag\\
&=v\int_{-\infty}^{W_2}dr_2'\int_{0}^{c_{12}r_2'}dr_1'\int_{-\infty}^{W_1+r_1'+\frac{1}{2}} dr_1 \int_{-\infty}^{r_2'+\frac{1}{2}} dr_2\,\tilde{\rho}_0(\bm{r})\notag\\
&\simeq v\int_{-\infty}^{+\infty} dr_1\int_{-\infty}^{+\infty}dr_2 \int_{r_2-\frac{1}{2}}^{W_2} dr_2'\int_{0}^{c_{12}r_2'}dr_1'\,\tilde{\rho}_0(\bm{r})\notag\\
&=\left\langle\int_{r_2-\frac{1}{2}}^{W_2} dr_2'\int_{0}^{c_{12}r_2'}dr_1'\,1\right\rangle_{\tilde{0}}\notag\\
&=\frac{1}{2}c_{12}\left\langle (W_2)^2-(r_2-\tfrac{1}{2})^2\right\rangle_{\tilde{0}}.
\end{align}
Interchanging the superscripts $1\leftrightarrow2$, we obtain
\begin{align}
Q_B-Q_E=\frac{1}{2}c_{21}\left\langle (W_1)^2-(r_1-\tfrac{1}{2})^2\right\rangle_{\tilde{0}}.
\end{align}
Finally, the charge neutrality in the bulk implies
\begin{align}
Q_C=0.
\end{align}
Plugging these expressions into Eq.~\eqref{QRQRp}, we find
\begin{align}
Q_R&=-W_1\langle r_2\rangle_0-W_2\langle r_1\rangle_0+\langle (r_1-\tfrac{1}{2})(r_2-\tfrac{1}{2})\rangle_{\tilde{0}}\notag\\
&\quad\quad-\frac{1}{2}c_{21}\langle (r_1-\tfrac{1}{2})^2\rangle_{\tilde{0}}-\frac{1}{2}c_{12}\langle (r_2-\tfrac{1}{2})^2\rangle_{\tilde{0}}.
\end{align}
When $\langle r_1\rangle_0\neq0$ or $\langle r_2\rangle_0\neq0$, this reproduces Eq.~\eqref{surfacecharge}. When $\langle r_1\rangle_0=\langle r_2\rangle_0=0$, we find
\begin{align}
Q_R=\langle Q_{12}(\bm{r})\rangle_{\tilde{0}}=\langle Q_{12}(\bm{r})\rangle_0,
\end{align}
verifying Eq.~\eqref{cornercharge}.

\subsection{Derivation of the formula under rotation symmetry}
\label{sec:quantization}

Now we move on to the derivation of Eq.~\eqref{rotationformula}. Our task is to properly impose the rotation symmetry on $\rho_0(\bm{r})$.  Note that the local charge density $\rho_0(\bm{r})$ itself is not necessarily $C_n$-symmetric, while the total charge density $\rho_{\text{tot}}^{\text{(bulk)}}(\bm{r})$ in Eq.~\eqref{totalcharge} must be $C_n$-symmetric:
\begin{equation}
\rho_{\text{tot}}^{\text{(bulk)}}(R_n^{-1}\bm{r})=\rho_{\text{tot}}^{\text{(bulk)}}(\bm{r}).
\label{rotbulk}
\end{equation}
Here and hereafter, we write the orthogonal matrix representing the $m$-fold rotation as
\begin{equation}
R_m=
\begin{pmatrix}
\cos\phi_m & -\sin\phi_m \\
\sin\phi_m & \cos\phi_m
\end{pmatrix},\quad\phi_m\equiv\frac{2\pi}{m}.
\end{equation}
for $m\in\mathbb{N}$.

To systematically study $\rho_0(\bm{r})$ that properly encodes the symmetry requirement, suppose that there is an orbital centered at a position $\bm{r}=\bm{r}_w^{(1)}$ with unit charge $e$.  When the site symmetry of the position $\bm{r}_w^{(1)}$ is $m_w$-fold rotation around $\bm{r}=\bm{r}_w^{(1)}$ ($m_w$ must be a divisor of $n$), the orbital must also be symmetric under $m_w$-fold rotation. Its contribution to $\rho_0(\bm{r})$ can be written as
\begin{equation}
e p_w(\bm{r}-\bm{r}_w^{(1)}),
\end{equation}
where  $p_w(\bm{r})$ is a $C_{m_w}$-symmetric unit density satisfying
\begin{align}
p_{m_w}(R_{m_w}^{-1}\bm{r})=p_w(\bm{r}),\quad\int_{\mathbb{R}^2} d^2rp_{m_w}(\bm{r})=1.
\label{pmsym}
\end{align}
The rotation invariance of $\rho_{\text{tot}}^{\text{(bulk)}}(\bm{r})$ in Eq.~\eqref{rotbulk} requires that a $(C_n)^{\ell-1}$-rotation copy of the orbital at $\bm{r}=\bm{r}_w^{(1)}$ must be placed at $\bm{r}=\bm{r}_w^{(\ell)}$ for $\ell=2,\dots,\nu_w$ ($\nu_w\equiv n/m_w$), where
\begin{align}
&\bm{r}_w^{(\ell)}-R_n\bm{r}_w^{(\ell-1)}=k_1^{(\ell)}\bm{a}_1+k_2^{(\ell)}\bm{a}_2
\label{condition1}
\end{align}
for some integers $k_{1}^{(\ell)}$ and $k_{2}^{(\ell)}$. These orbitals altogether give contribution
\begin{align}
\rho_{0w}(\bm{r})\equiv e\sum_{\ell=1}^{\nu_w}p_w(R_n^{1-\ell}(\bm{r}-\bm{r}_w^{(\ell)}))
\end{align}
to $\rho_0(\bm{r})$. Their contribution to $Q_c$ can be written as
\begin{align}
\langle Q_{12}(\bm{r})\rangle_{0w}&\equiv\int_{\mathbb{R}^2} d^2r\rho_{0w}(\bm{r})Q_{12}(\bm{r})\notag\\
&=e\int_{\mathbb{R}^2} d^2rp_w(\bm{r})\sum_{\ell=1}^{\nu_w}Q_{12}(R_n^{\ell-1}\bm{r}+\bm{r}_w^{(\ell)}).
\label{Qrw}
\end{align}

\begin{figure}
\begin{center}
\includegraphics[width=1\columnwidth]{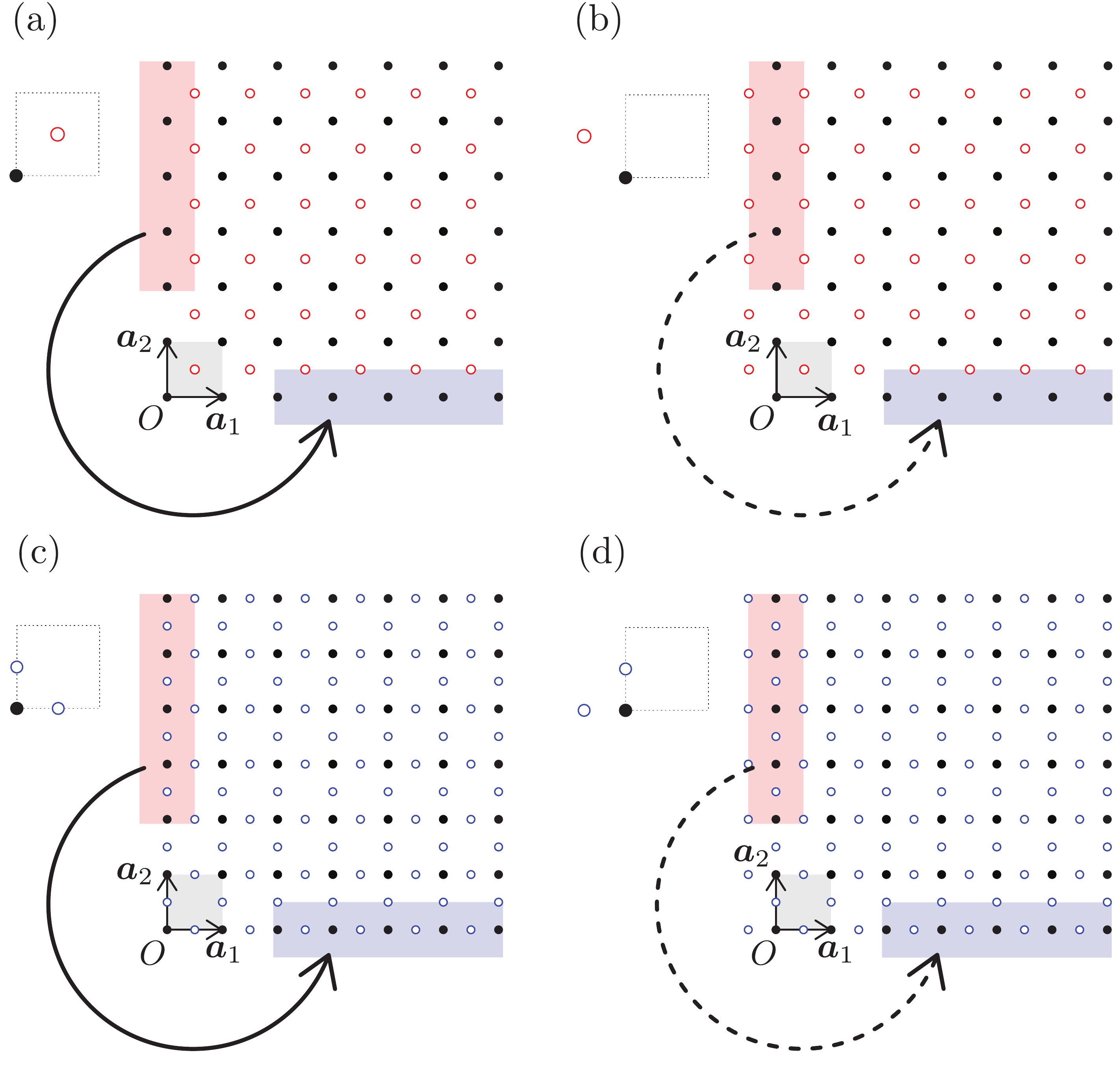}
\caption{\label{figC42} The four-fold rotation symmetry between the surface along $\bm{a}_1$ (blue) and the surface along $\bm{a}_2$ (red). The symmetry is respected in panel (a) and (c), and is violated in (b) and (d). The repetition unit $\rho_0(\bm{r})$ is shown at the bottom-left of each panel. }
\end{center}
\end{figure}

The condition \eqref{condition1} was imposed for the rotation invariance of the bulk charge density.
We additionally require that the two surfaces normal to $\bm{b}_1$ and $\bm{b}_2$ must be related to each other by the $n$-fold rotation symmetry.  
For example, in Fig.~\ref{figC42}, both panels (a) and (b) have the same bulk charge distribution, but the $C_4$ symmetry between the two surfaces is violated in (b).  

The symmetry of the surfaces can be implemented in the following way.  On the one hand, the charge density at the surface along $\bm{a}_1$ for $r_1\gg \lambda$ is given by
\begin{align}
\rho_{\text{tot}}^{\text{(1)}}(\bm{r})=\sum_{n_1\in\mathbb{Z}}\sum_{n_2\geq0}\rho_0(\bm{r}-n_1\bm{a}_1-n_2\bm{a}_2).
\end{align}
On the other hand, the charge density at the surface along $\bm{a}_2$ for $r_2\gg \lambda$ is
\begin{align}
\rho_{\text{tot}}^{\text{(2)}}(\bm{r})=\sum_{n_1\geq0}\sum_{n_2\in\mathbb{Z}}\rho_0(\bm{r}-n_1\bm{a}_1-n_2\bm{a}_2).
\end{align}
If rotated by an angle $\phi_n=2\pi/n$ and shifted along $\bm{a}_1$, $\rho_{\text{tot}}^{\text{(2)}}(\bm{r})$ should coincide with $\rho_{\text{tot}}^{\text{(1)}}(\bm{r})$:
\begin{align}
&\rho_{\text{tot}}^{\text{(2)}}(R_n^{-1}(\bm{r}-m\bm{a}_1))=\rho_{\text{tot}}^{\text{(1)}}(\bm{r}).
\end{align}
Using $R_n\bm{a}_1=\bm{a}_2+2\cos\phi_n\bm{a}_1$ and $R_n\bm{a}_2=-\bm{a}_1$ (recall our choice of $\theta=\pi-\phi_n$), this condition can be rewritten as
\begin{align}
&\sum_{n_1\in\mathbb{Z}}\sum_{n_2\geq0}\rho_0(R_n^{-1}(\bm{r}-n_1\bm{a}_1-n_2\bm{a}_2))\notag\\
&=\sum_{n_1\in\mathbb{Z}}\sum_{n_2\geq0}\rho_0(\bm{r}-n_1\bm{a}_1-n_2\bm{a}_2),
\end{align}
which means that the rotated pattern $\rho_0(R_n^{-1}\bm{r})$ and the original pattern $\rho_0(\bm{r})$ give the same charge distribution when translated along $\bm{a}_1$. This requirement imposes additional constraints on $k_i^{(\ell)}$ in Eq.~\eqref{condition1}:
\begin{align}
&k_2^{(\ell)}=0\quad (\ell=2,\ldots,\nu_w)
\label{condition2}
\end{align}
and
\begin{align}
&\bm{r}_w^{(1)}-R_n\bm{r}_w^{(\nu_w)}=k_1^{(1)}\bm{a}_1
\label{condition3}
\end{align}
for an integer $k_1^{(1)}$. 

Using the rotation symmetry of $p_w(\bm{r})$ in Eq.~\eqref{pmsym} and conditions on $k_i^{(\ell)}$ in Eqs.~\eqref{condition2} and \eqref{condition3}, we find that $\langle Q_{12}(\bm{r})\rangle_{0w}$ in Eq.~\eqref{Qrw} becomes
\begin{align}
\langle Q_{12}(\bm{r})\rangle_{0w}=e\sum_{\ell=1}^{\nu_w}Q_{12}(\bm{r}_w^{(\ell)}).
\end{align}
That is, the contribution to $\langle Q_{12}(\bm{r})\rangle_{0}$ from a unit charge $e$ placed at $\bm{r}_w^{(\ell)}$ does not depend on the detailed shape of the orbital $p_w(\bm{r})$.

The classification of distinct $\bm{r}_w^{(\ell)}$'s, satisfying Eqs.~\eqref{condition1}, \eqref{condition2}, and \eqref{condition3}, is related to the notion of Wyckoff positions~\cite{ITC}.  In the following, we summarize the possible choices of  $\bm{r}_w^{(1)}$ for $n=4$, $6$, and $3$ one by one.

So far we separately studied the contribution from orbitals generated from $\bm{r}_w^{(1)}$. These building blocks must be carefully superposed in order to form $\rho_0(\bm{r})$ that satisfied the charge neutrality and has vanishing bulk polarization.  Examples of valid superpositions can be found in Figs.~\ref{figC4} and \ref{figC6C3}.

\subsubsection{$C_4$}
For $C_4$ symmetry, we set $a_1=a_2=a$ and $\theta=\pi/2$. In this case $Q_{12}(\bm{r})$ in Eq.~\eqref{cornercharge} reduces to
\begin{align}
&Q_{12}(\bm{r})=r_1r_2=\frac{x y}{a^2}.
\label{C4formula}
\end{align}
As the choice of $\bm{r}_w^{(\ell)}$', we have
\begin{align}
&\bm{r}_a^{(1)}=k(\bm{a}_1+\bm{a}_2),
\label{4a}
\end{align}
\begin{align}
&\bm{r}_b^{(1)}=(\tfrac{1}{2}+k)(\bm{a}_1+\bm{a}_2),
\end{align}
\begin{align}
&\bm{r}_c^{(1)}=(\tfrac{1}{2}+k)\bm{a}_1+k'\bm{a}_2,\notag\\
&\bm{r}_c^{(2)}=k'\bm{a}_1+(\tfrac{1}{2}+k)\bm{a}_2,
\label{4c}
\end{align}
\begin{align}
&\bm{r}_d^{(1)}=r_1\bm{a}_1+r_2\bm{a}_2,\notag\\
&\bm{r}_d^{(2)}=(k-r_2)\bm{a}_1+r_1\bm{a}_2\notag\\
&\bm{r}_d^{(3)}=(k'-r_1)\bm{a}_1+(k-r_2)\bm{a}_2,\notag\\
&\bm{r}_d^{(4)}=r_2\bm{a}_1+(k'-r_1)\bm{a}_2.
\label{4d}
\end{align}
In these expressions, integers $k$, $k'$ (related to $k_1^{(\ell)}$) can be set freely.  The standard convention of the Wyckoff position in Ref.~\onlinecite{ITC} is recovered by setting $k=k'=0$.
We tabulate the value of $\langle P_1(\bm{r})\rangle_0=\langle P_2(\bm{r})\rangle_0$ and $\langle Q_{12}(\bm{r})\rangle_0$ originating from a unit charge placed at these positions in Table~\ref{tabC4}. From this table, we see that $\langle Q_{12}(\bm{r})\rangle_{0}=q_b/4$.

To generate a charge-neutral and polarization-free insulator, the U(1) charge per site $q_w$ at each Wyckoff position  must satisfy
\begin{align}
&\sum_w\nu_w q_w=q_a+q_b+2q_c+4q_d=0,\\
&q_c=q_b\quad\mod 2e,
\end{align}
which imply
\begin{align}
q_a=-3q_b=q_b\mod 4e.
\end{align}
Thus we obtain Eq.~\eqref{rotationformula} for $n=4$.

\begin{table}[t]
  \begin{center}
    \caption{The contribution to the bulk polarization $\langle P_i(\bm{r})\rangle_0$ and the quadrupole $\langle Q_{12}(\bm{r})\rangle_0$ from a unit charge $e$ placed at the position $\bm{r}_w^{(\ell)}$ ($\ell=1,\dots,\nu_w$) for $C_4$-invariant systems. See Eqs.~\eqref{4a}-\eqref{4d} for the definition of $\bm{r}_w^{(\ell)}$.\label{tabC4}}
    \begin{tabular}{c|cccc} \hline\hline
$w$ & $m_w$ & $\nu_w$ & $\langle P_i(\bm{r})\rangle_{0w}/e$ & $\langle Q_{12}(\bm{r})\rangle_{0w}/e$ (mod 1)\\\hline
$a$ & 4 & $1$ & $k$ & $0$ \\
$b$ & 4 & $1$ & $\frac{1}{2}+k$ & $\frac{1}{4}$ \\
$c$ & 2 & $2$ & $\frac{1}{2}+k+k'$ & $0$ \\
$d$ & 1 & $4$ & $k+k'$ & $0$ \\ \hline\hline
    \end{tabular}
  \end{center}
\end{table}

\subsubsection{$C_6$}
For $n=6$, we set $a_1=a_2=a$ and $\theta=2\pi/3$. In this case, Eq.~\eqref{cornercharge} reads
\begin{align}
&Q_{12}(\bm{r})\equiv r_1r_2-\frac{1}{4}(r_1^2+r_2^2)=\frac{y^2+2\sqrt{3}xy-x^2}{4a^2}.
\end{align}
We have
\begin{align}
&\bm{r}_a=k(\bm{a}_1+\bm{a}_2),
\label{6a}
\end{align}
\begin{align}
&\bm{r}_b^{(1)}=(\tfrac{2}{3}+k)\bm{a}_1+(\tfrac{1}{3}+k')\bm{a}_2,\notag\\
&\bm{r}_b^{(2)}=(\tfrac{1}{3}+k')\bm{a}_1+(\tfrac{2}{3}+k)\bm{a}_2,
\end{align}
\begin{align}
&\bm{r}_c^{(1)}=(\tfrac{1}{2}+k)\bm{a}_1+k'\bm{a}_2,\notag\\
&\bm{r}_c^{(2)}=(\tfrac{1}{2}+k-k'+k'')\bm{a}_1+(\tfrac{1}{2}+k)\bm{a}_2,\notag\\
&\bm{r}_c^{(3)}=k'\bm{a}_1+(\tfrac{1}{2}+k-k'+k'')\bm{a}_2,
\end{align}
and
\begin{align}
&\bm{r}_d^{(1)}=r_1\bm{a}_1+r_2\bm{a}_2,\notag\\
&\bm{r}_d^{(2)}=(r_1-r_2+k)\bm{a}_1+r_1\bm{a}_2,\notag\\
&\bm{r}_d^{(3)}=-(r_2-k-k')\bm{a}_1+(r_1-r_2+k)\bm{a}_2,\notag\\
&\bm{r}_d^{(4)}=-(r_1-k'-k'')\bm{a}_1-(r_2-k-k')\bm{a}_2,\notag\\
&\bm{r}_d^{(5)}=(r_2-r_1-k+k''+k''')\bm{a}_1-(r_1-k'-k'')\bm{a}_2,\notag\\
&\bm{r}_d^{(6)}=r_2\bm{a}_1+(r_2-r_1-k+k''+k''')\bm{a}_2
\label{6d}
\end{align}
for integers $k$, $k'$, $k''$, $k'''$.  We tabulate the value of $\langle P_1(\bm{r})\rangle_0=\langle P_2(\bm{r})\rangle_0$ and $\langle Q_{12}(\bm{r})\rangle_0$ originating from a unit charge placed at these positions in Table~\ref{tabC6}.

To prove Eq.~\eqref{rotationformula}, let us define vectors $\vec{w}\equiv(\nu_w,\langle P_i(\bm{r})\rangle_{0w}/e,\langle Q_{12}(\bm{r})\rangle_{0w}/e)$ mod $(0,0,1)$ for each $w$:
\begin{align}
&\vec{a}(k)=(1,k,\tfrac{1}{2}k),\\
&\vec{b}(k,k')=(2,1+k+k',\tfrac{1}{6}+\tfrac{1}{2}k+\tfrac{1}{2}k'),\\
&\vec{c}(k,k',k'')=(3,1+2k+k'',\tfrac{1}{2}k''),\\
&\vec{d}(k,k',k'',k''')=(6,k+2k'+2k''+k''',\tfrac{1}{2}k+\tfrac{1}{2}k''').
\end{align}
Subtracting $\vec{a}(k)$ to annihilate the first and second entry, we find
\begin{align}
\vec{b}'&\equiv\vec{b}(k,k')-\vec{a}(k)-\vec{a}(k'+1)=(0,0,\tfrac{2}{3}),\\
\vec{c}'&\equiv\vec{c}(k,k',k'')-2\vec{a}(k)-\vec{a}(k''+1)=(0,0,\tfrac{1}{2}),\\
\vec{d}'&\equiv\vec{d}(k,k',k'',k''')-\vec{a}(k)-2\vec{a}(k')-2\vec{a}(k'')-\vec{a}(k''')\notag\\
&=(0,0,0).
\end{align}
Any charge-neutral and polarization-free insulator can be decomposed into a superposition of $\vec{b}'$, $\vec{c}'$, and $\vec{d}'$, and their occupation coincides with $q_b$, $q_c$, and $q_d$:
\begin{align}
q_b\vec{b}'+q_c\vec{c}'+q_d\vec{d}'=(0,0,\tfrac{2}{3}q_b+\tfrac{1}{2}q_c)
\end{align}
Hence, we find $\langle Q_{12}(\bm{r})\rangle_{0}=(2/3)q_b+(1/2)q_c$ modulo $e$.  

Finally, the charge neutrality implies that
\begin{align}
\sum_w\nu_w q_w=q_a+2q_b+3q_c+6q_d=0.
\end{align}
Thus, 
\begin{align}
q_a=-2q_b-3q_c=4q_b+3q_c\mod 6e.
\end{align}
These relations prove Eq.~\eqref{rotationformula} for $n=6$.

\begin{table}
\begin{center}
\caption{The contribution to the bulk polarization $\langle P_i(\bm{r})\rangle_0$ and the quadrupole $\langle Q_{12}(\bm{r})\rangle_0$ from a unit charge $e$ placed at the position $\bm{r}_w^{(\ell)}$ ($\ell=1,\dots,\nu_w$) for $C_6$-invariant systems.  See Eqs.~\eqref{6a}-\eqref{6d} for the definition of $\bm{r}_w^{(\ell)}$.\label{tabC6}}
\begin{tabular}{c|cccc} \hline\hline
$w$ & $m_w$ & $\nu_w$ & $\langle P_i(\bm{r})\rangle_{0w}/e$ & $\langle Q_{12}(\bm{r})\rangle_{0w}/e$ (mod 1)\\\hline
$a$ & 6 & $1$ & $k$ & $\frac{1}{2}k$ \\
$b$ & 3 & $2$ & $1+k+k'$ & $\frac{1}{6}+\frac{1}{2}k+\frac{1}{2}k'$ \\
$c$ & 2 &  $3$ & $1+2k+k''$ & $\frac{1}{2}k''$ \\
$d$ & 1 & $6$ & $k+2k'+2k''+k'''$ & $\frac{1}{2}k+\frac{1}{2}k'''$ \\ \hline\hline
\end{tabular}
\end{center}
\end{table}

\subsubsection{$C_3$}
For $n=3$, we set $a_1=a_2=a$ and $\theta=\pi/3$.
In this case, Eq.~\eqref{cornercharge} becomes
\begin{align}
&Q_{12}(\bm{r})\equiv r_1r_2+\frac{1}{4}(r_1^2+r_2^2)=\frac{x^2+2\sqrt{3}xy-y^2}{4a^2}.
\end{align}
We have
\begin{align}
&\bm{r}_a=k(\bm{a}_1+\bm{a}_2),
\label{3a}
\end{align}
\begin{align}
&\bm{r}_b=(\tfrac{1}{3}+k)(\bm{a}_1+\bm{a}_2),
\end{align}
\begin{align}
&\bm{r}_c=(\tfrac{2}{3}+k)(\bm{a}_1+\bm{a}_2),
\end{align}
and
\begin{align}
&\bm{r}_c^{(1)}=r_1\bm{a}_1+r_2\bm{a}_2,\notag\\
&\bm{r}_c^{(2)}=-(r_1+r_2-k)\bm{a}_1+r_1\bm{a}_2,\notag\\
&\bm{r}_c^{(3)}=r_2\bm{a}_1-(r_1+r_2-k)\bm{a}_2
\label{3d}
\end{align}
for an integer $k$. We tabulate the value of $\langle P_1(\bm{r})\rangle_0=\langle P_2(\bm{r})\rangle_0$ and $\langle Q_{12}(\bm{r})\rangle_0$ originating from a unit charge placed at these positions in Table~\ref{tabC3}.

\begin{table}
  \begin{center}
    \caption{The contribution to the bulk polarization $\langle P_i(\bm{r})\rangle_0$ and the quadrupole $\langle Q_{12}(\bm{r})\rangle_0$ from a unit charge $e$ placed at the position $\bm{r}_w^{(\ell)}$ ($\ell=1,\dots,\nu_w$) for $C_3$-invariant systems.  See Eqs.~\eqref{3a}-\eqref{3d} for the definition of $\bm{r}_w^{(\ell)}$.\label{tabC3}}
    \begin{tabular}{c|cccc} \hline\hline
$w$ &  $m_w$ & $\nu_w$ & $\langle P_i(\bm{r})\rangle_{0w}/e$ & $\langle Q_{12}(\bm{r})\rangle_{0w}/e$ (mod 1)\\\hline
$a$ & $3$ &$1$ & $k$ & $\frac{1}{2}k$ \\
$b$ & $3$ & $1$ & $\tfrac{1}{3}+k$ & $\tfrac{1}{6}+\frac{1}{2}k$ \\
$c$ & $3$ & $1$ & $\tfrac{2}{3}+k$ & $\tfrac{2}{3}+\frac{1}{2}k$ \\
$d$ & $1$ & $3$ & $k$ & $\frac{1}{2}k$ \\ \hline\hline
    \end{tabular}
  \end{center}
\end{table}

To generate a charge-neutral and polarization-free insulator, the U(1) charge $q_w$ at the Wyckoff position $\bm{r}_w^{(\ell)}$ must satisfy
\begin{align}
&\sum_w\nu_w q_w=q_a+q_b+q_c+3q_d=0,\\
&q_c=q_b\quad\mod 3e,
\end{align}
which implies
\begin{align}
q_a=-2q_b=q_b\mod 3e.
\end{align}

We define vectors $\vec{w}\equiv(\nu_w,\langle P_i(\bm{r})\rangle_{0w}/e,\langle Q_{12}(\bm{r})\rangle_{0w}/e)$ mod $(0,0,1)$ for each $w$ as before:
\begin{align}
&\vec{a}(k)=(1,k,\tfrac{1}{2}k),\\
&\vec{b}(k)=(1,\tfrac{1}{3}+k,\tfrac{1}{6}+\tfrac{1}{2}k),\\
&\vec{c}(k)=(1,\tfrac{2}{3}+k,\tfrac{2}{3}+\tfrac{1}{2}k),\\
&\vec{d}(k)=(3,k,\tfrac{1}{2}k).
\end{align}
Subtracting $\vec{a}(k)$ to partially annihilate the first and second entry, we find
\begin{align}
\vec{a}'&\equiv\vec{a}(k)-\vec{a}(k-1)=(0,1,\tfrac{1}{2}),\\
\vec{b}'&\equiv\vec{b}(k)-\vec{a}(k+1)=(0,-\tfrac{2}{3},\tfrac{2}{3}),\\
\vec{c}'&\equiv\vec{c}(k)-\vec{a}(k)=(0,\tfrac{2}{3},\tfrac{2}{3}),\\
\vec{d}'&\equiv\vec{d}(k)-\vec{a}(k)-2\vec{a}(0)=(0,0,0).
\end{align}
Taking superposition of these vectors with coefficients $q_a'$, $q_b$, $q_c$, and $q_d$, we get
\begin{align}
&q_a'\vec{a}'+q_b\vec{b}'+q_c\vec{c}'+q_d\vec{d}'\notag\\
&=(0,q_a'-\tfrac{2}{3}(q_b-q_c),\tfrac{1}{2}q_a'+\tfrac{2}{3}(q_b+q_c))\notag\\
&=(0,0,\tfrac{1}{3}q_b)\mod (0,0,e).
\end{align}
In the last step we set $q_a'/e=(2/3e)(q_b-q_c)\in\mathbb{Z}$. Thus Eq.~\eqref{rotationformula} for $n=3$ is verified.

\section{Three dimensions}
\label{threedim}
Let us generalize discussions above to three-dimensional systems.  We will see that basically the same calculation applies.
Our results are Eqs.~\eqref{hingecharge} and \eqref{cornercharge3D} that give the charge density localized to the hinge and the corner in terms of the bulk quadrupole moment and the bulk octupole moment, respectively.

Extending Eq.~\eqref{chargedensity} to $d=3$, the coarse-grained total charge density at position  $\bm{r}=r_1\bm{a}_1+r_2\bm{a}_2+r_3\bm{a}_3$ can be expressed as
\begin{align}
\tilde{\rho}_{\text{tot}}(\bm{r})=\int_{-\infty}^{r_1+\frac{1}{2}}dr_1'\int_{-\infty}^{r_2+\frac{1}{2}}dr_2'\int_{-\infty}^{r_3+\frac{1}{2}}dr_3'\tilde{\rho}_0(\bm{r}').
\end{align}
The total charge in the region $R$, defined by Eq.~\eqref{RR}, is thus given by
\begin{widetext}
\begin{align}
Q_R\equiv \int_{R}d^3r\tilde{\rho}_{\text{tot}}(\bm{r})=v\int_{-\infty}^{W_1+c_{12}r_2+c_{13}r_3}dr_1\int_{-\infty}^{W_2+c_{21}r_1+c_{23}r_3}dr_2\int_{-\infty}^{W_3+c_{31}r_1+c_{32}r_2}dr_3\tilde{\rho}_{\text{tot}}(\bm{r}).
\end{align}
Here $c_{ij}$ is defined in Eq.~\eqref{cij}. We decompose this integral into eight distinct pieces.
\begin{align}
&v\left(\int_{-\infty}^{W_1}dr_1+\int_{W_1}^{W_1+c_{12}r_2+c_{13}r_3}dr_1\right)
\left(\int_{-\infty}^{W_2}dr_2+\int_{W_2}^{W_2+c_{21}r_1+c_{23}r_3}dr_2\right)
\left(\int_{-\infty}^{W_3}dr_3+\int_{W_3}^{W_3+c_{31}r_1+c_{32}r_2}dr_3\right)\tilde{\rho}_0(\bm{r})\notag\\
&=Q_{R'}+Q_{R_{1}}+Q_{R_{2}}+Q_{R_{3}}+Q_{R_{23}}+Q_{R_{31}}+Q_{R_{12}}+Q_{R_{123}},
\label{3Ddecomposition}
\end{align}
\end{widetext}
where
\begin{align}
Q_{R'}&\equiv v\int_{-\infty}^{W_1}dr_1\int_{-\infty}^{W_2}dr_2\int_{-\infty}^{W_3}dr_3\tilde{\rho}_{\text{tot}}(\bm{r}),
\end{align}
\begin{align}
Q_{R_{3}}&\equiv v\int_{-\infty}^{W_1}dr_1\int_{-\infty}^{W_2}dr_2\int_{W_3}^{W_3+c_{31}r_1+c_{32}r_2}dr_3\tilde{\rho}_{\text{tot}}(\bm{r}),
\end{align}
\begin{align}
Q_{R_{12}}&\equiv v\int_{W_1}^{W_1+c_{12}r_2+c_{13}r_3}dr_1\notag\\
&\quad\quad\quad\int_{W_2}^{W_2+c_{21}r_1+c_{22}r_2}dr_2\int_{-\infty}^{W_3}dr_3\tilde{\rho}_{\text{tot}}(\bm{r}),
\end{align}
and
\begin{align}
Q_{R_{123}}&\equiv v\int_{W_1}^{W_1+c_{12}r_2+c_{13}r_3}dr_1\int_{W_2}^{W_2+c_{21}r_1+c_{23}r_3}dr_2\notag\\
&\quad\quad\quad\quad\quad\quad\quad\int_{W_3}^{W_3+c_{31}r_1+c_{32}r_2}dr_3\tilde{\rho}_{\text{tot}}(\bm{r}).
\end{align}
Other components of $Q_{R_{i}}$ and $Q_{R_{ij}}$ are defined similarly. Their concrete expressions can be generated by interchanging the superscript $1\rightarrow 2\rightarrow 3\rightarrow 1$.  The decomposition in Eq.~\eqref{3Ddecomposition} is the analog of Eq.~\eqref{QRQRp} for two-dimensional systems.  

In the same way as in Sec.~\ref{sec:derivation}, we find
\begin{align}
Q_{R'}&=\left\langle \int_{r_1-\frac{1}{2}}^{W_1}dr_1'\int_{r_2-\frac{1}{2}}^{W_2}dr_2'\int_{r_3-\frac{1}{2}}^{W_3}dr_3'\,1\right\rangle_{\tilde{0}}\notag\\
&= \big\langle  \left(W_1-(r_1-\tfrac{1}{2})\right) \notag\\
&\quad\quad\times\left(W_2-(r_2-\tfrac{1}{2})\right)\left(W_3-(r_3-\tfrac{1}{2})\right)\big\rangle_{\tilde{0}},
\end{align}
\begin{align}
Q_{R_{3}}&=\left\langle \int_{r_1-\frac{1}{2}}^{W_1}dr_1'\int_{r_2-\frac{1}{2}}^{W_2}dr_2'\int_0^{c_{31}r_1'+c_{32}r_2'}dr_3'\,1\right\rangle_{\tilde{0}}\notag\\
&= \frac{1}{2}c_{31}\left\langle \left(W_1^2-(r_1-\tfrac{1}{2})^2\right)\left(W_2-(r_2-\tfrac{1}{2})\right)\right\rangle_{\tilde{0}}\notag\\
&+ \frac{1}{2}c_{32} \left\langle  \left(W_2^2-(r_2-\tfrac{1}{2})^2\right) \left(W_1-(r_1-\tfrac{1}{2})\right)\right\rangle_{\tilde{0}}.
\end{align}
We also have
\begin{align}
Q_{R_{123}}=0
\end{align}
because of the charge neutrality in the bulk.

\begin{figure}
\begin{center}
\includegraphics[width=\columnwidth]{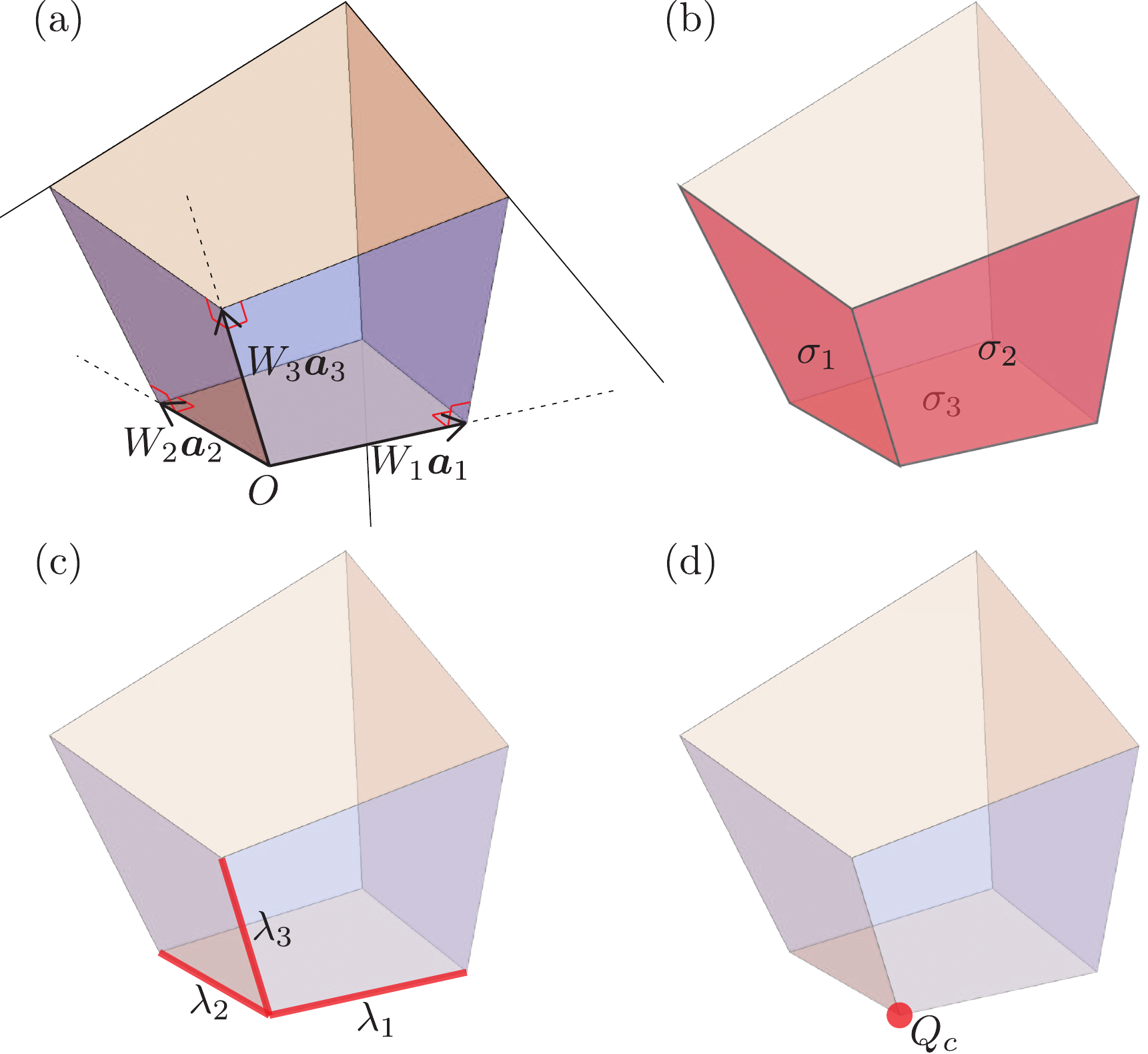}
\caption{\label{fig3D} Three dimensional version of Fig.~\ref{figR}. The illustration of (a) the region $R$, (b) the surface charge density $\sigma_i$, (c) the hinge charge density $\lambda_i$,  and (d) the corner charge $Q_c$.}
\end{center}
\end{figure}

It remains to evaluate $Q_{R_{12}}$. This term can be expressed as
\begin{align}
Q_{R_{12}}=\left\langle \int_{r_3-\frac{1}{2}}^{W_3}dr_3'A(r_3')\right\rangle_{\tilde{0}},
\end{align}
where
\begin{align}
A(r_3')&\equiv\int_0^{c_{12}(r_2'+W_2)+c_{13}r_3'}dr_1'\int_0^{c_{21}(r_1'+W_1)+c_{23}r_3'}dr_2'\,1
\end{align}
is the area surrounded by four lines $r_1'=r_2'=0$, $r_1'=c_{12}(r_2'+W_2)+c_{13}r_3'$, and $r_2'=c_{21}(r_1'+W_1)+c_{23}r_3'$. Using the property $c_{12} c_{21}=\frac{(\bm{a}_1\cdot\bm{a}_2)^2}{a_1^2a_2^2}<1$, we find
\begin{align}
A(r_3')&=\frac{c_{12}c_{21}(c_{21} W_1^2 +2  W_1 W_2+c_{12} W_2^2)}{2(1-c_{12} c_{21})}\notag\\
&\quad+\frac{c_{21}(c_{13}+c_{12}c_{23})W_1+c_{12}(c_{23} + c_{21}c_{13})W_2}{1-c_{12} c_{21}}r_3'\notag\\
&\quad+\frac{c_{12}(c_{23})^2+2 c_{13}c_{23}+c_{21}(c_{13})^2}{2(1-c_{12} c_{21})}(r_3')^2.
\end{align}
Therefore,
\begin{widetext}
\begin{align}
Q_{R_{12}}&=  \frac{c_{12}c_{21}(c_{21} W_1^2 +2  W_1 W_2+c_{12} W_2^2)}{2(1-c_{12} c_{21})}\left\langle W_3-(r_3-\tfrac{1}{2})\right\rangle_{\tilde{0}}\notag\\
&\quad+\frac{c_{21}(c_{13}+c_{12}c_{23})W_1+c_{12}(c_{23} + c_{21}c_{13})W_2}{2(1-c_{12} c_{21})}\left\langle W_3^2-(r_3-\tfrac{1}{2})^2\right\rangle_{\tilde{0}}\notag\\
&\quad+\frac{c_{12}(c_{23})^2+2 c_{13}c_{23}+c_{21}(c_{13})^2}{6(1-c_{12} c_{21})}\left\langle W_3^3-(r_3-\tfrac{1}{2})^3\right\rangle_{\tilde{0}}.
\end{align}
\end{widetext}
Plugging all of these expressions into Eq.~\eqref{3Ddecomposition}, we find
\begin{align}
Q_R=\sum_{i=1}^3S_i\sigma_i+\sum_{i=1}^3W_i\lambda_i+Q_c,
\end{align}
where 
\begin{align}
S_3&\equiv W_1W_2+\frac{1}{2}\left(c_{21}W_1^2+c_{12}W_2^2\right)\notag\\
&\quad+\frac{c_{12}c_{21}(c_{21} W_1^2 +2  W_1 W_2+c_{12} W_2^2)}{2(1-c_{12} c_{21})}
\end{align}
is the area of the surface normal to $\bm{b}_3$ [the part marked $\sigma_3$ in Fig.~\ref{fig3D} (b)] divided by $|\bm{a}_1\times\bm{a}_2|$. $S_1$ and $S_2$ are defined in the same manner.

The surface charge density $\sigma_3$ per unit area $|\bm{a}_1\times\bm{a}_2|$ [Fig.~\ref{fig3D} (b)] is given by the bulk polarization
\begin{align}
\sigma_3&=-\langle P_3(\bm{r})\rangle_0=-\langle P_3(\bm{r})\rangle_{\tilde{0}},\\
P_i(\bm{r})&\equiv\bm{b}_i\cdot\bm{r}=r_i.
\end{align}
Again, the electric contribution to the bulk polarization can be expressed as the Berry phase of filled bands [see Eq.~\eqref{Berry} below].

When the surface charge density $\sigma_1$ and $\sigma_2$ vanish ($\sigma_3$ can be nonzero), the hinge charge density $\lambda_3$  [Fig.~\ref{fig3D} (c)] becomes well defined.  It is given by a bulk quadrupole moment:
\begin{align}
\lambda_{3}&=\langle Q_{12}(\bm{r})\rangle_0=\langle Q_{12}(\bm{r})\rangle_{\tilde{0}}\\
Q_{12}(\bm{r})&\equiv(\bm{b}_1\cdot\bm{r})(\bm{b}_2\cdot\bm{r})+\frac{\bm{a}_1\cdot\bm{b}_1\times\bm{a}_3}{2\bm{a}_2\cdot\bm{b}_1\times\bm{a}_3}(\bm{b}_1\cdot\bm{r})^2\notag\\
&\quad+\frac{\bm{a}_2\cdot\bm{b}_2\times\bm{a}_3}{2\bm{a}_1\cdot\bm{b}_2\times\bm{a}_3}(\bm{b}_2\cdot\bm{r})^2\notag\\
&=r_1r_2-\frac{1}{2}\left(c_{21}+c_{23}\frac{c_{31}+c_{32}c_{21}}{1-c_{23}c_{32}}\right)(r_1)^2\notag\\
&\quad-\frac{1}{2}\left(c_{12}+c_{13}\frac{c_{32}+c_{31}c_{12}}{1-c_{31}c_{13}}\right)(r_2)^2.\label{hingecharge} 
\end{align}
The expression for $\lambda_{1}=\langle O_{23}(\bm{r})\rangle_0$ and $\lambda_{2}=\langle O_{31}(\bm{r})\rangle_0$ can be found by interchanging the superscript $1\rightarrow 2\rightarrow 3\rightarrow 1$.

Finally, when the surface charge density $\sigma_i$ and the hinge charge density $\lambda_i$ all vanish, the corner charge $Q_c$  [Fig.~\ref{fig3D} (d)] becomes well defined. It is given by the bulk octupole moment:
\begin{align}
&Q_c=-\langle O_{123}(\bm{r})\rangle_0=-\langle O_{123}(\bm{r})\rangle_{\tilde{0}},\\
&O_{123}(\bm{r})= r_1r_2r_3-\frac{1}{2}[c_{23} (r_3)^2+c_{32} (r_2)^2]r_1\notag\\
&-\frac{1}{2}[c_{31} (r_1)^2+c_{13} (r_3)^2]r_2-\frac{1}{2}[c_{12} (r_2)^2+c_{21} (r_1)^2]r_3\notag\\
&\quad+\frac{c_{23}(c_{31})^2+2 c_{21}c_{31}+c_{32}(c_{21})^2}{6(1-c_{23} c_{32})}(r_1)^3\notag\\
&\quad+\frac{c_{31}(c_{12})^2+2 c_{32}c_{12}+c_{13}(c_{32})^2}{6(1-c_{31} c_{13})}(r_2)^3\notag\\
&\quad+\frac{c_{12}(c_{23})^2+2 c_{13}c_{23}+c_{21}(c_{13})^2}{6(1-c_{12} c_{21})}(r_3)^3.\label{cornercharge3D} 
\end{align}
We find that all of $P_i(\bm{r})$, $Q_{ij}(\bm{r})$, and $O_{123}(\bm{r})$ possess the property of the coarse-graining invariance in Eq.~\eqref{invariance}.

One should be able to discuss the quantization of $\langle Q_{ij}(\bm{r})\rangle_0$ and $\langle O_{123}(\bm{r})\rangle_0$ in the presence of point-group symmetry and derive their formulas in terms of Wyckoff positions. We will leave the comprehensive analysis of these important problems as future work. Here we instead discuss a single example of the cubic system with the space group symmetry $P432$ (No.~207).

For this space group, there are 11 Wyckoff positions $\bm{r}_w^{(\ell)}$ ($w=a,b,\dots,k$) in total (see Ref.~\onlinecite{ITC} for details). Four of them are with a maximal site symmetry:
\begin{eqnarray}
\bm{r}_a^{(1)}=(0,0,0),
\end{eqnarray}
\begin{eqnarray}
\bm{r}_b^{(1)}=(\tfrac{1}{2},\tfrac{1}{2},\tfrac{1}{2}),
\end{eqnarray}
\begin{eqnarray}
&\bm{r}_c^{(1)}=(0,\tfrac{1}{2},\tfrac{1}{2}),\quad \bm{r}_c^{(2)}=(\tfrac{1}{2},0,\tfrac{1}{2}),\notag\\
&\bm{r}_c^{(3)}=(\tfrac{1}{2},\tfrac{1}{2},0).
\end{eqnarray}
\begin{eqnarray}
&\bm{r}_d^{(1)}=(\tfrac{1}{2},0,0),\quad\bm{r}_d^{(2)}=(0,\tfrac{1}{2},0),\notag\\
&\bm{r}_d^{(3)}=(0,0,\tfrac{1}{2}).
\end{eqnarray}
Other seven Wyckoff positions have some free parameters. For example,
\begin{eqnarray}
&\bm{r}_f^{(1)}=(\xi,\tfrac{1}{2},\tfrac{1}{2}),\quad \bm{r}_f^{(2)}=(-\xi,\tfrac{1}{2},\tfrac{1}{2}),\notag\\
&\bm{r}_f^{(3)}=(\tfrac{1}{2},\xi,\tfrac{1}{2}),\quad \bm{r}_f^{(4)}=(\tfrac{1}{2},-\xi,\tfrac{1}{2}),\notag\\
&\bm{r}_f^{(5)}=(\tfrac{1}{2},\tfrac{1}{2},\xi),\quad \bm{r}_f^{(6)}=(\tfrac{1}{2},\tfrac{1}{2},-\xi).
\label{wpf}
\end{eqnarray}
Note that $\bm{r}_f^{(\ell)}$'s reduce to six copies of $\bm{r}_b^{(1)}$ by setting $\xi\rightarrow1/2$, and to two sets of $\bm{r}_c^{(\ell)}$'s by setting $\xi\rightarrow0$. This implies that $q_b$ is defined only modulo $6e$.

In fact, by the same analysis as in Sec.~\ref{sec:quantization}, we find
\begin{eqnarray}
\langle O_{123}(\bm{r})\rangle_{\tilde{0}}=\frac{1}{8}q_a=\frac{1}{8}q_b\mod \frac{1}{4}e.
\end{eqnarray}
The $e/4$-ambiguity follows immediately from the $6e$ ambiguity of $q_b$. The same conclusion can also be reached from the perspective of surface decoration.
If the system is symmetrically decorated, as illustrated in Fig.~\ref{figdecoration} (b), with two-dimensional $C_4$-symmetric quadrupole insulators with a quantized corner charge $ne/4$, then the corner charge of the three-dimensional system $Q_c$ is changed by $3ne/4$. This implies that $Q_c$ is well defined only modulo $e/4$.  
We learn an important lesson from this example: The corner charge is affected by surface decoration more severely in three dimensions and even the fractional part can be altered.
A recent study~\cite{NaCl} revealed that sodium chloride, one of the most popular crystals around us, is actually an example of octupole insulators with fractional corner charge $\pm e/8$.

\label{sec:3D}

\section{Discussions}
\label{sec:discussion}
In this work, we developed a framework of describing the hinge charge density and the corner charge in terms of the bulk quadrupole moment and octupole moment.
We derived expressions in Eqs.~\eqref{cornercharge},  \eqref{hingecharge}, and \eqref{cornercharge3D} for the particular type of corners and hinges formed by planes normal to $\bm{b}_i$ ($i=1,\dots,d$). We also discussed the rotation symmetric cases for two-dimensional systems and proved that the fractional part of the corner charge can be predicted solely from the bulk point of view using the formula in Eq.~\eqref{rotationformula}. In three dimensions, we focused on a cubic system with $P432$ symmetry and revealed that the corner charge $Q_c$ has $e/4$ ambiguity.

In our formalism electrons and ions are treated on the same footing. One immediate implication is that when the sign of $e$ is flipped, so is the sign of $Q_c$. 
Recently, Ref.~\onlinecite{Hirayama_honeycomb} discussed a $C_6$-symmetric insulator in which an ion sits at the honeycomb site (the Wyckoff position $w=b$) and two electrons are at the triangular lattice site (the Wyckoff position $w=a$). For this example, $q_b=+e$ and $q_a=-2e$ and we immediately gets $Q_c=2q_b/3=2e/3$ mod $e$ from Eq.~\eqref{rotationformula}. This is well anticipated in our formalism because the insulator in which electrons and ions are interchanged [see Fig.~\ref{figC6C3} (c)] is known to have $Q_c=e/3=-2e/3$ mod $e$~\cite{benalcazar2018}.

In the presence of both spin-orbit coupling and the time-reversal symmetry, all electronic orbitals must form Kramers' pairs. If the same is true for all ions in the system, the charge unit is effectively doubled and one can replace $e$ with $e'=2e$ in all formulas derived in this work. As an example, let us discuss the $C_4$-symmetric insulator discussed in Ref.~\onlinecite{PhysRevResearch.1.033074}, for which the rotation representations are silent for the corner charge. In this example, four electrons occupy the Wyckoff position $w=b$ ($q_b=-2e'=-4e$) and the same number of ions sit at the Wyckoff position $w=a$ ($q_a=+2e'=+4e$) so that Eq.~\eqref{rotationformula} predicts $Q_c=q_a/4=e'/2=e$ mod $e'=2e$. Thus, our formula goes beyond the formalism based on the rotation representations of the Bloch functions. More generally, a symmetry-indicator type approach utilizes only restricted information (i.e., representations of the little group) of the Bloch functions and does not have the full resolution on the topological nature of the band insulator. 

Note that the Kramers' doubling does not necessarily apply to all ions in the problem. That is, in principle, it is allowed to consider a time-reversal invariant cation with a charge $+e$ and an integer spin. The simplest example would be the deuterium ion (deuteron) $\text{D}^+={}^2\text{He}^+$ whose total angular momentum is $1$. If this type of ions is taken into account, the charge unit remains $e$ and the odd-integer corner charge can be annihilated by adding such a cation. 

In retrospect, Eq.~\eqref{rotationformula} has a simple interpretation associated to the filling anomaly~\cite{benalcazar2018}. If the bulk of a $C_n$-symmetric insulating phase is charge neutral and polarization free, Eq.~\eqref{rotationformula} suggests that the corner charge $Q_c$ is given by the U(1) charge $q$ bound to the $n$-fold rotation axis:
\begin{equation}
Q_c=\frac{1}{n}q\mod e.\label{Qq}
\end{equation}
On the other hand,  the corner charge $Q_c$ is also given by the total U(1) charge $Q$ in the entire system under a $C_n$-symmetric open boundary condition~\cite{benalcazar2018,PhysRevX.9.031003}:
\begin{equation}
Q_c=\frac{1}{n}Q\mod e.
\label{Qq222}
\end{equation}
These two expressions are consistent since the local charge $q$ and the total charge $Q$ can differ only by an integer multiple of $ne$ as indicated by Eq.~\eqref{Qwqw}, although they are conceptually quite distinct because $q$ in Eq.~\eqref{Qq} is a bulk quantity of the system while $Q$ in Eq.~\ref{Qq222} is a global quantity. These formulas can also be used to describe the corner charge of $C_n$-symmetric insulating phases \emph{without} translation symmetry for arbitrary $n\in\mathbb{N}$. Examples include quasi-crystals, disordered systems, and lattices with disclinations~\cite{benalcazar2018,PhysRevX.9.031003,PhysRevB.101.115115}.

\vspace{1.0\baselineskip}

\emph{Note added:}
Recently, we had email exchanges with the author of Ref.~\onlinecite{Luka2020}. According to the author, Eq.~(32) of his work, if $\vec{n}_\alpha'$'s and $\vec{e}_\alpha'$'s in it are properly chosen, corresponds to Eq.~(23) of ours.

\begin{acknowledgements}
We would like to thank Hoi Chun Po, Yohei Fuji, Ryosuke Hirakida, and Luka Trifunovic for useful discussions. 
This work was initiated in the workshop ``Recent Developments on Multipole Moments in Quantum Systems" partly organized by H.W. and we thank the active and constructive discussions by attendees.
In particular, we obtained several important clues from the oral presentation by David Vanderbilt. 
The work of S.O. is supported by Materials Education program for the future leaders in Research, Industry, and Technology (MERIT) and  KAKENHI Grant No.~JP20J21692 from the JSPS. 
The work of H.W. is supported by JSPS KAKENHI Grant No.~JP20H01825 and by JST PRESTO Grant No.~JPMJPR18LA. 
\end{acknowledgements}

\appendix
\section{Band insulators}
\label{sec:BI}
Here we derive the expression of $\rho_0(\bm{r})$ for band insulators.  Let us consider a tight-binding model
\begin{align}
\hat{H}&=\sum_{\bm{R}\bm{R}'\sigma\sigma'}\hat{c}_{\bm{R}\sigma}^\dagger (h_{\bm{R}'-\bm{R}})_{\sigma\sigma'}\hat{c}_{\bm{R}'\sigma'}.
\end{align}
Here, $\bm{R}=\sum_{i=1}^dn_i\bm{a}_i$ specifies a unit cell and $\hat{c}_{\bm{R}\sigma}^\dagger$ is the creation operator of an electron with the orbital label $\sigma$ at the position $\bm{r}=\bm{R}+\bm{x}_\sigma$. In this section, we assume the periodic boundary condition and denote by $N$ the total number of unit cells in the system. After the Fourier transformation
\begin{align}
&\hat{c}_{\bm{R}\sigma}^\dagger=\frac{1}{\sqrt{N}}\sum_{\bm{k}} \hat{c}_{\bm{k}\sigma}^\dagger e^{-i\bm{k}\cdot(\bm{R}+\bm{x}_\sigma)},\\
&(h_{\bm{R}'-\bm{R}})_{\sigma\sigma'}=\frac{1}{N}\sum_{\bm{k}} (h_{\bm{k}})_{\sigma\sigma'}e^{-i\bm{k}\cdot(\bm{R}'+\bm{x}_{\sigma'}-\bm{R}-\bm{x}_\sigma)},
\end{align}
the Hamiltonian becomes
\begin{align}
\hat{H}=\sum_{\bm{k}\sigma\sigma'} \hat{c}_{\bm{k}\sigma}^\dagger (h_{\bm{k}})_{\sigma\sigma'}\hat{c}_{\bm{k}\sigma}=\sum_{n\bm{k}} \hat{\gamma}_{n\bm{k}}^\dagger \epsilon_{n\bm{k}}\hat{\gamma}_{n\bm{k}},
\end{align}
where
\begin{align}
\hat{\gamma}_{n\bm{k}}^\dagger\equiv\sum_\sigma\hat{c}_{\bm{k}\sigma}^\dagger u_{n\bm{k}\sigma}
\end{align}
is the creation operator of a Bloch electron in the $n$-th band and $u_{n\bm{k}\sigma}$ is an eigenvector of $(h_{\bm{k}})_{\sigma\sigma'}$ normalized as $\sum_\sigma u_{n\bm{k}\sigma}^*u_{n'\bm{k}\sigma}=\delta_{nn'}$.  The insulating ground state can be expressed as
\begin{align}
|\Phi\rangle\equiv\prod_{n\in\text{occ}}\prod_{\bm{k}}\hat{\gamma}_{n\bm{k}}^\dagger |0\rangle.
\end{align}
The electronic contribution of the total charge density is given by
\begin{align}
&\rho_{\text{tot}}^{\text{(el)}}(\bm{r})=-e\sum_{\bm{R}\sigma}\langle\Phi|\hat{c}_{\bm{R}\sigma}^\dagger\hat{c}_{\bm{R}\sigma}|\Phi\rangle\delta^d(\bm{r}-\bm{R}-\bm{x}_\sigma)\notag\\
&=-\frac{e}{N}\sum_{\bm{R}\sigma}\sum_{n\in\text{occ}}\sum_{\bm{k}}|u_{n\bm{k}\sigma}|^2\delta^d(\bm{r}-\bm{R}-\bm{x}_\sigma).
\end{align}
This quantity is gauge-invariant; i.e., is independent of the choice of the phase of $u_{n\bm{k}}$.

Let us switch to the Wannier basis by a unitary transformation 
\begin{align}
\hat{w}_{n\bm{R}_0}^\dagger&\equiv\frac{1}{\sqrt{N}}\sum_{\bm{k}}\hat{\gamma}_{n\bm{k}}^\dagger e^{-i\bm{k}\cdot\bm{R}_0}\notag\\
&=\frac{1}{\sqrt{N}}\sum_{\bm{R}\sigma}\hat{c}_{\bm{R}\sigma}^\dagger w_{n\sigma}(\bm{R}-\bm{R}_0),
\end{align}
where
\begin{align}
w_{n\sigma}(\bm{R})\equiv\frac{1}{\sqrt{N}}\sum_{\bm{k}}u_{n\bm{k}\sigma}e^{i\bm{k}\cdot(\bm{R}+\bm{x}_\sigma)}
\end{align}
is the Wannier orbital belonging to the unit cell $\bm{R}_0=\bm{0}$.  In this basis, the ground state $|\Phi_0\rangle$ and the total charge density $\rho_{\text{tot}}^{\text{(el)}}(\bm{r})$ can be written as $|\Phi_0\rangle=\prod_{n\in\text{occ}}\prod_{\bm{R}_0}\hat{w}_{n\bm{R}_0}^\dagger |0\rangle$ and 
\begin{align}
&\rho_{\text{tot}}^{\text{(el)}}(\bm{r})\notag\\
&=-\frac{e}{N}\sum_{\bm{R}\sigma}\sum_{\bm{R}_0}\sum_{n\in\text{occ}}|w_{n\sigma}(\bm{R}-\bm{R}_0)|^2\delta^d(\bm{r}-\bm{R}-\bm{x}_\sigma)\notag\\
&=-\frac{e}{N}\sum_{\bm{R}_0}\sum_{\bm{R}\sigma}\sum_{n\in\text{occ}}|w_{n\sigma}(\bm{R})|^2\delta^d(\bm{r}-\bm{R}_0-\bm{R}-\bm{x}_\sigma).
\end{align}
Therefore, we identify
\begin{align}
&\rho_0^{\text{(el)}}(\bm{r})\equiv-\frac{e}{N}\sum_{\bm{R}\sigma}\sum_{n\in\text{occ}}|w_{n\sigma}(\bm{R})|^2\delta^d(\bm{r}-\bm{R}-\bm{x}_\sigma)\notag\\
&=-\frac{e}{N^2}\sum_{\bm{R}\sigma}\sum_{n\in\text{occ}}\sum_{\bm{k}\bm{k}'}u_{n\bm{k}\sigma}^*u_{n\bm{k}'\sigma}e^{i(\bm{k}'-\bm{k})\cdot\bm{r}}
\delta^d(\bm{r}-\bm{R}-\bm{x}_\sigma).
\label{rho0wannie}
\end{align}
This quantity depends on the choice of the phase of the Bloch function $u_{n\bm{k}\sigma}$~\cite{marzari1997,RevModPhys.84.1419}. The electric polarization is given by
\begin{align}
&\langle P_i(\bm{r})\rangle_0^{\text{(el)}}\equiv\int d^dr \rho_0^{\text{(el)}}(\bm{r})r_i\notag\\
&=-\frac{e}{N}\sum_{\bm{R}\sigma}\sum_{n\in\text{occ}}|w_{n\sigma}(\bm{R})|^2\bm{b}_i\cdot(\bm{R}+\bm{x}_\sigma)\notag\\
&=-\frac{ie}{N}\sum_{\bm{k}\sigma}\sum_{n\in\text{occ}} u_{n\bm{k}\sigma}^*(\bm{b}_i\cdot\bm{\nabla}_{\bm{k}})u_{n\bm{k}\sigma}.
\label{Berry}
\end{align}
The gauge transformation $u_{n\bm{k}\sigma}\rightarrow e^{2\pi i\theta_{n\bm{k}}}u_{n\bm{k}\sigma}$ changes $\langle P_i(\bm{r})\rangle_0^{\text{(el)}}/e$ by an integer amount  $\theta_{n\bm{k}+2\pi\bm{b}_i}-\theta_{n\bm{k}}$.

By analogy, it is tempting to express $\langle Q_{ij}(\bm{r})\rangle_0^{\text{(el)}}$ using the proper combination of 
\begin{align}
&\int d^dr \rho_0^{\text{(el)}}(\bm{r})r_ir_j\notag\\
&=-\frac{e}{N}\sum_{\bm{R}\sigma}\sum_{n\in\text{occ}}|w_{n\sigma}(\bm{R})|^2\bm{b}_i\cdot(\bm{R}+\bm{x}_\sigma)\bm{b}_j\cdot(\bm{R}+\bm{x}_\sigma)\notag\\
&=\frac{e}{N}\sum_{\bm{k}\sigma}\sum_{n\in\text{occ}}u_{n\bm{k}\sigma}^*(\bm{b}_i\cdot\bm{\nabla}_{\bm{k}})(\bm{b}_j\cdot\bm{\nabla}_{\bm{k}})u_{n\bm{k}\sigma}.
\end{align}
However, computing $\langle Q_{ij}(\bm{r})\rangle_0^{\text{(el)}}$ from this type of expression is dangerous, because it is difficult to properly implement the assumed rotation symmetry  in the corresponding $\rho_0^{\text{(el)}}(\bm{r})$ in Eq.~\eqref{rho0wannie}.  In practice, it is much easier and safer to use Eq.~\eqref{rotationformula} instead.

\bibliography{ref}

\end{document}